\documentclass{article}

\usepackage{arxiv}

\usepackage[utf8]{inputenc}
\usepackage{tocloft,amsmath,amsfonts,amssymb,amsthm,mathtools,dsfont,bbold,bbm,graphicx,fancybox,enumitem,hyperref,comment,xcolor,caption, subcaption, booktabs}

\def\first#1{\fcolorbox{black}{black!15}{\textbf{#1}}}
\def\second#1{\fcolorbox{black!50}{black!5}{#1}}

\let\OldXi\Xi
\def\Xi{\mathit \OldXi}
\def\RR{\mathbb{R}}

\title{Testing a QUBO Formulation of Core-periphery Partitioning on a  Quantum Annealer}
\author{
    \textbf{Catherine F. Higham} \\
    School of Computing Science\\
    University of Glasgow\\
     Glasgow G12 8QQ, UK \\
    \texttt{catherine.higham@glasgow.ac.uk} 
    \and
    \textbf{Desmond J. Higham} \\
    School of Mathematics\\
    University of Edinburgh\\
    Edinburgh EH93FD, UK\\
    \texttt{d.j.higham@ed.ac.uk} 
    \and
    \textbf{Francesco Tudisco}\\
  School of Mathematics\\
  Gran Sasso Science Institute\\
  67100, L'Aquila, Italy \\
  \texttt{francesco.tudisco@gssi.it} 
}

\begin{document}

\maketitle

\begin{abstract}
    We propose a new kernel that quantifies success 
    for the task of 
    computing a core-periphery partition for an 
    undirected network. Finding the associated optimal partitioning
     may be expressed in the form of a 
     quadratic unconstrained 
    binary optimization (QUBO) problem, to which  
    a state-of-the-art quantum annealer may be applied.
    We therefore make use of the new objective function to 
    (a) judge the performance of a quantum annealer, 
    and (b) compare this approach with existing heuristic 
    core-periphery
    partitioning methods.
    The quantum annealing is performed on the commercially available D-Wave machine.
    The QUBO problem involves a full matrix even when the underlying network is sparse. Hence, we develop and test a 
    sparsified version of the original QUBO which increases the 
    available problem dimension for the quantum annealer.
    Results 
     are provided on both synthetic and real data sets, and we conclude    that the QUBO/quantum annealing approach
     offers benefits in terms of optimizing this new quantity of interest.
\end{abstract}

\section{Motivation}
\label{sec:mot}

Clustering, or community detection,
is a fundamental tool for
extracting 
high-level information 
from a network \cite{EK15}.
However, it is now widely acknowledged that 
quantifying and discovering 
other forms of meso-scale structure may also reveal 
useful insights.
In this work we look at the issue of identifying core–periphery structure; we seek a set of nodes that are highly connected 
both internally and with the rest of the network, forming the core, and a set of peripheral nodes that are well connected to the core but have only sparse internal connections.
This type of core-periphery
structure has been observed to arise naturally 
in a number of settings, including 
protein interaction, cell signalling, gene regulation,
ecology, social interaction and global trade; see, for example, 
\cite{csermely2013structure} for a review.
Further, as pointed out in 
\cite{BK18}, the structure may arise 
as a consequence of the data collection process. 
For example,    
a phone service provider may only have access to calls in which at least one of the participants is a customer; so there will be no record of 
calls between pairs of non-customers, who thus inhabit the periphery.
We are concerned in this work with the ``inverse problem'' where a 
set of nodes and (undirected, unweighted) edges are supplied, and the task is to partition the nodes into a core and periphery; this may provide useful information about the roles of individual nodes and may
also lead to more instructive visualizations 
\cite{borgatti2000models,csermely2013structure,ZTM15}.

A second motivation for this work is the 
recent development in quantum annealing, which 
has the potential to 
outperform classical methodologies on certain 
classes of discrete optimization problem
\cite{CL21}.
In particular, the company 
D-Wave  
(\texttt{dwavesys.com})
offers direct commercial access to a quantum annealer.

The main contributions of this work are
\begin{itemize}
    \item to develop a kernel-based objective function 
    that quantifies success in the problem of 
    discovering a core-periphery node partition,
    \item to exploit the fact that this leads naturally to a quadratic unconstrained 
    binary optimization (QUBO) problem, and to study 
    how a state-of-the-art quantum annealer performs in this context,
    \item to use the objective function to compare existing heuristic algorithms 
    \item for problem dimensions 
    small enough to allow the quantum annealer to be used,
     to compare the output from 
     heuristic algorithms with the 
     quantum annealed ``global'' optimum,
     \item for larger problem dimensions associated with sparse networks, to show that a nearby sparse QUBO problem can give good results.
\end{itemize}

\section{Related Work}
\label{sec:rel}

The concept of a network core-periphery structure was formalized
and studied by Borgatti and Everett \cite{borgatti2000models}.
As mentioned in this work, and also noted by many 
subsequent authors \cite{csermely2013structure,cucuringu2016detection,GYW21}, there are several different types of 
core-periphery structure, and hence 
detection algorithm, that can be defined.
First, we may distinguish between 
\emph{partitions}
\cite{brusco11,FR20,ZTM15}
that 
map nodes
into two sets, the core and the periphery, and 
\emph{orderings}
\cite{cucuringu2016detection,kitsak2010identification,rombach2017core,tudisco2021nonlinear} that assign a nonnegative ``coreness'' score to each node. 
The latter 
are closely associated with node centrality measures \cite{holme2005core}, and, of course,
a continuous score can be used for subsequent ranking and partitioning. 
Second, while there is general consistency around the principle
that core nodes should be well-connected and peripheral nodes 
should be poorly connected, there is a choice to be made about whether edges that 
join a core node and a peripheral node should occur with 
high/intermediate frequency \cite{cucuringu2016detection,P21,tudisco2019core,ZTM15} or low frequency \cite{GYW21}, or whether such edges are irrelevant
\cite{brusco11,FR20,holme2005core}.

In this work, we 
focus on the partitioning task and we 
take the view that 
core-periphery connections should 
occur with high or intermediate frequency.
The adjacency matrix plots in Figure~\ref{fig:rhos}
illustrate this type of ``L-shaped'' two-by-two block structure.
In common with
\cite{borgatti2000models,cucuringu2016detection,rombach2017core,tudisco2019core}
we define an objective function that 
measures the extent to which a partition reveals 
a core-periphery structure, and we consider the resulting 
discrete optimization problem.
Our focus is on designing a well-motivated and simple objective
function that 
is parameter-free and does not require the core and periphery
size to be predefined. We also show that our optimization problem has QUBO form and hence 
is amenable to quantum annealing, giving us the opportunity
to compare 
results from existing partitioning algorithms 
with the ``global'' optimum (modulo imperfections in the
physical annealing process).

\section{Optimization Formulation}
\label{sec:opt}

\subsection{Notation}
\label{subsec:notation}

For our given undirected, unweighted network of $N$ nodes with no self-loops, we let $A \in \RR^{N \times N}$ denote the adjacency matrix; so 
$a_{ij} = 1$ if nodes $i$ and $j$ share an edge and 
$a_{ij} = 0$ otherwise.
We also let $D \in \RR^{N \times N}$ be the diagonal degree matrix with 
$d_{ii} = \mathrm{deg}_i = \sum_{j=1}^{N} a_{ij}$.
We use
$\mathbf{1} \in \RR^N$ to denote the vector with all elements
equal to one, 
$I \in \RR^{N \times N}$ to denote the identity matrix,  
and $E = \mathbf{1} \mathbf{1}^T - I \in \RR^{N \times N} $ 
to denote the adjacency matrix for the complete graph.

\subsection{Objective Function}
\label{subsec:objective}

We will use 
$x \in \RR^{N}$
as the indicator vector for a core-periphery partition, with the convention that 
$x_i =1$ assigns node $i$ to the core and 
$x_i =0$ assigns node $i$ to the periphery.
A useful starting point, adopted 
by several authors, see for example, 
\cite[equation~(1)]{borgatti2000models} and 
\cite[subsection~4.2.1]{rombach2017core},
is to consider maximizing 
over all choices of $x_i \in \{0,1\}$ the objective function
\begin{equation}
\sum_{i=1}^{N} \sum_{j=1}^{N} a_{ij} \max\{x_i,x_j\}.
\label{eq:maxobj}
\end{equation}
A motivation for 
(\ref{eq:maxobj}) is that we get one added to the sum every time
we have an edge $a_{ij} =1$ involving at least one core node.
However, directly 
maximizing (\ref{eq:maxobj}) 
is not practical, since 
the obvious solution is to assign every node to the core.
Hence, we must add constraints or alter the objective function.

One criticism of 
(\ref{eq:maxobj}) is that it 
does not take account of the missing 
edges, which should arise between periphery-periphery pairs.
This motivates the maximization of 
\begin{equation}
\sum_{i=1}^{N} \sum_{j=1, j \neq i}^{N}
a_{ij}
 \max\{x_i,x_j\} + (1-a_{ij}) ( 1 -  \max\{x_i,x_j\}).
\label{eq:obj1}
\end{equation}
In this objective function 
we get one added to the sum every time
we have an edge involving at least one core node \textbf{and} 
 every time
we have a missing edge involving no core nodes.

However, (\ref{eq:obj1}) suffers from a drawback when the
network is sparse. Here, the objective function 
(\ref{eq:obj1})
encourages the placement of all nodes into the 
periphery---in this way all missing edges 
contribute positively to the sum since they involve
periphery-periphery pairs.
Similarly, for a dense network, (\ref{eq:obj1}) 
encourages the placement of all nodes into the core.
For this reason, it makes sense to scale the two terms 
in (\ref{eq:obj1}) in relation to the numbers of edges that are present and missing. We therefore consider
maximizing 
\begin{equation}
\sum_{i=1}^{N} \sum_{j=1, j \neq i}^{N}
a_{ij} 
 \max\{x_i,x_j\} 
 \frac{1}{N_1}
 + (1-a_{ij}) ( 1 -  \max\{x_i,x_j\}) \frac{1}{N_2},
\label{eq:obj1b}
\end{equation}
where $N_1$ and $N_2$ denote the number of present and missing edges, respectively;  so $N_1 + N_2 = N (N-1)/2$.
Intuitively, up to a constant factor $N_1 + N_2$, we can interpret
(\ref{eq:obj1b}) as
dividing the count for ``good edges'' by $N_1/(N_1+N_2)$ (which is the probability that we see an edge if we choose a
pair of nodes at random)
and also  
dividing the count for ``good missing edges'' by $N_2/(N_1+N_2)$ (which is the probability that we see a missing 
edge if we choose a
pair of nodes at random).
Hence, we take a weighted combination of the 
number of correct edges and correct missing 
edges arising from the partition $x$, accounting for 
the relative probabilities of seeing each type.

We note that in contrast to previously defined objective functions, 
there are no user-defined parameters 
in (\ref{eq:obj1b}) and there is no requirement to 
specify the core size ahead of time.

\begin{figure}[htbp]
\centering
\scalebox{0.3}{\includegraphics{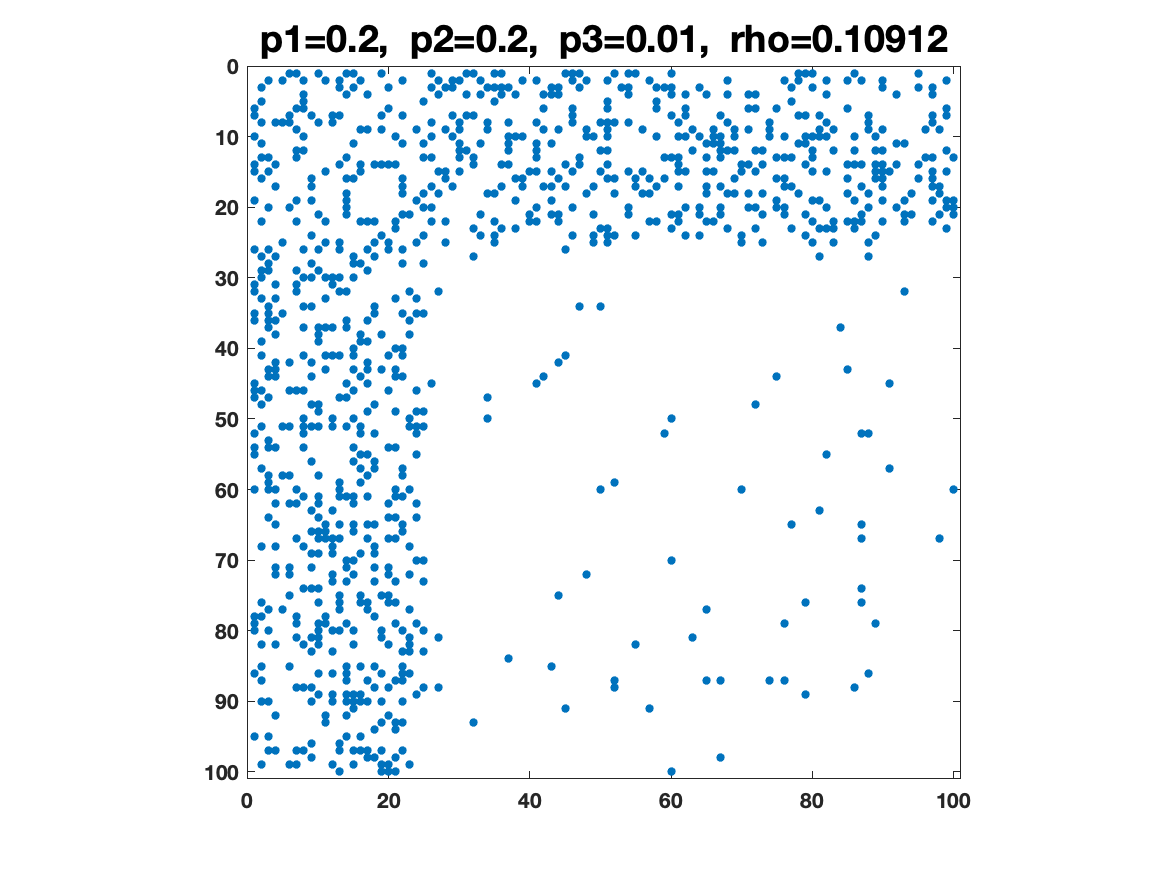}}
\scalebox{0.3}{\includegraphics{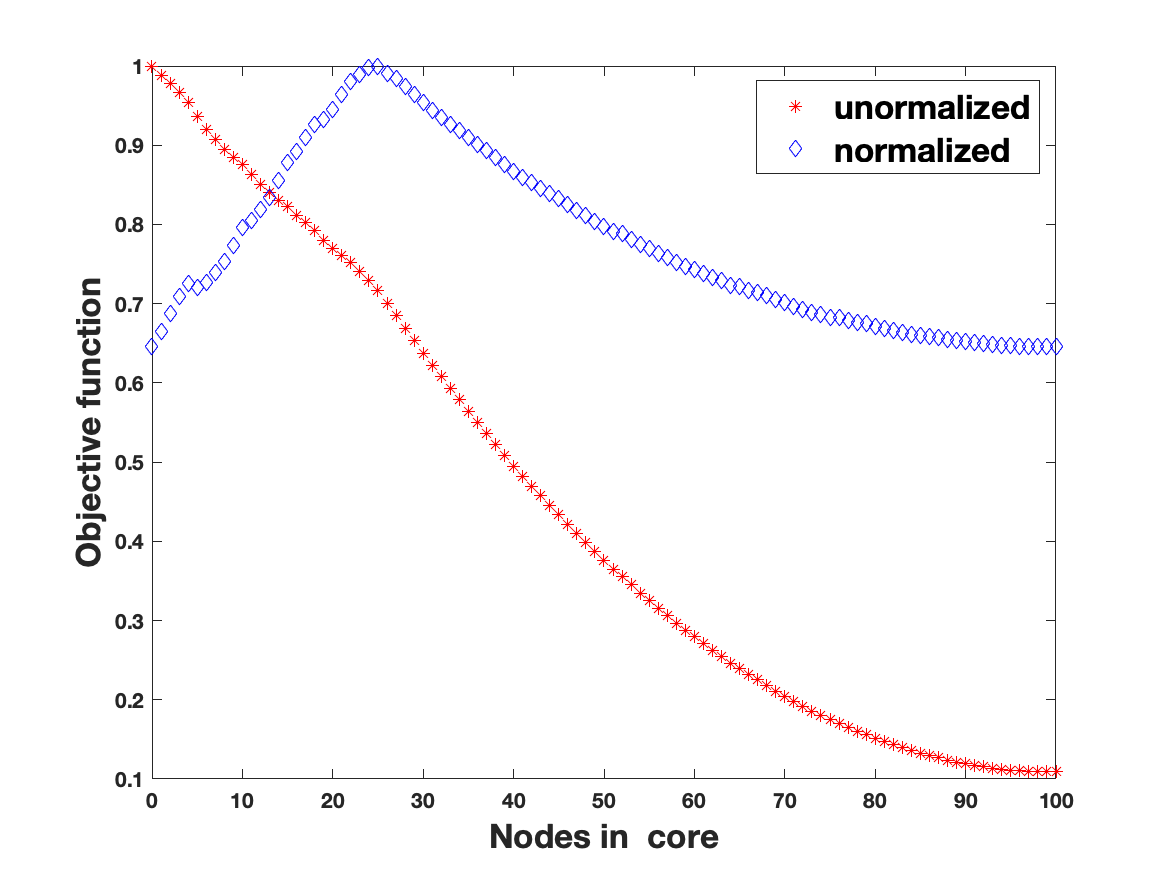}}
\scalebox{0.3}{\includegraphics{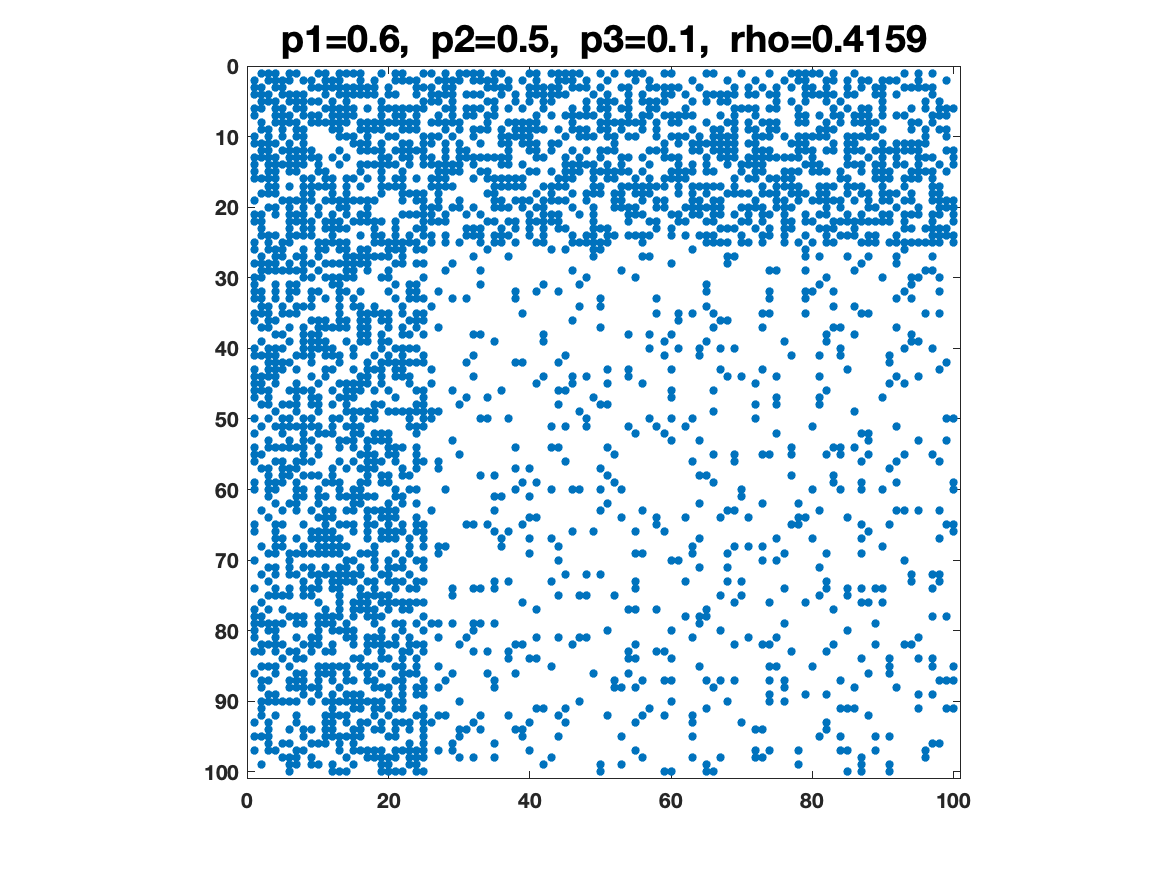}}
\scalebox{0.3}{\includegraphics{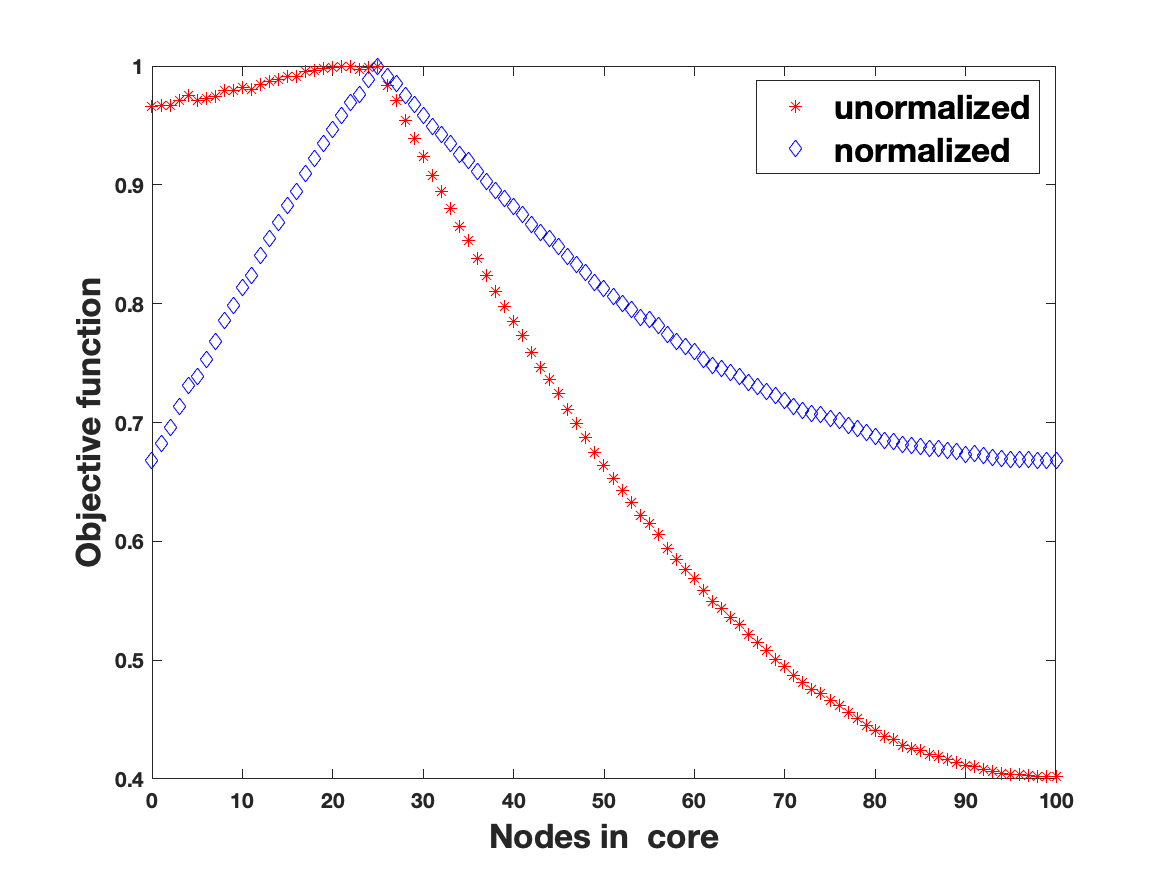}}
\scalebox{0.3}{\includegraphics{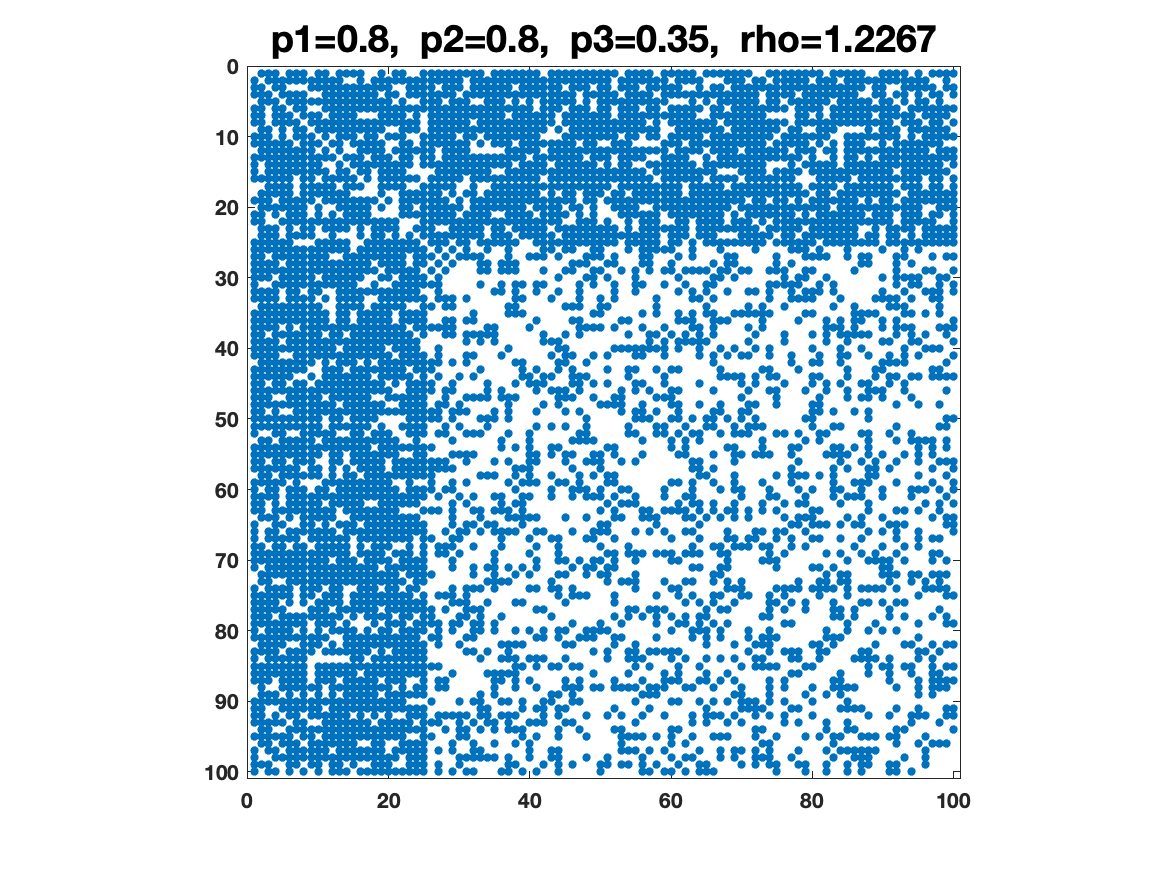}}
\scalebox{0.3}{\includegraphics{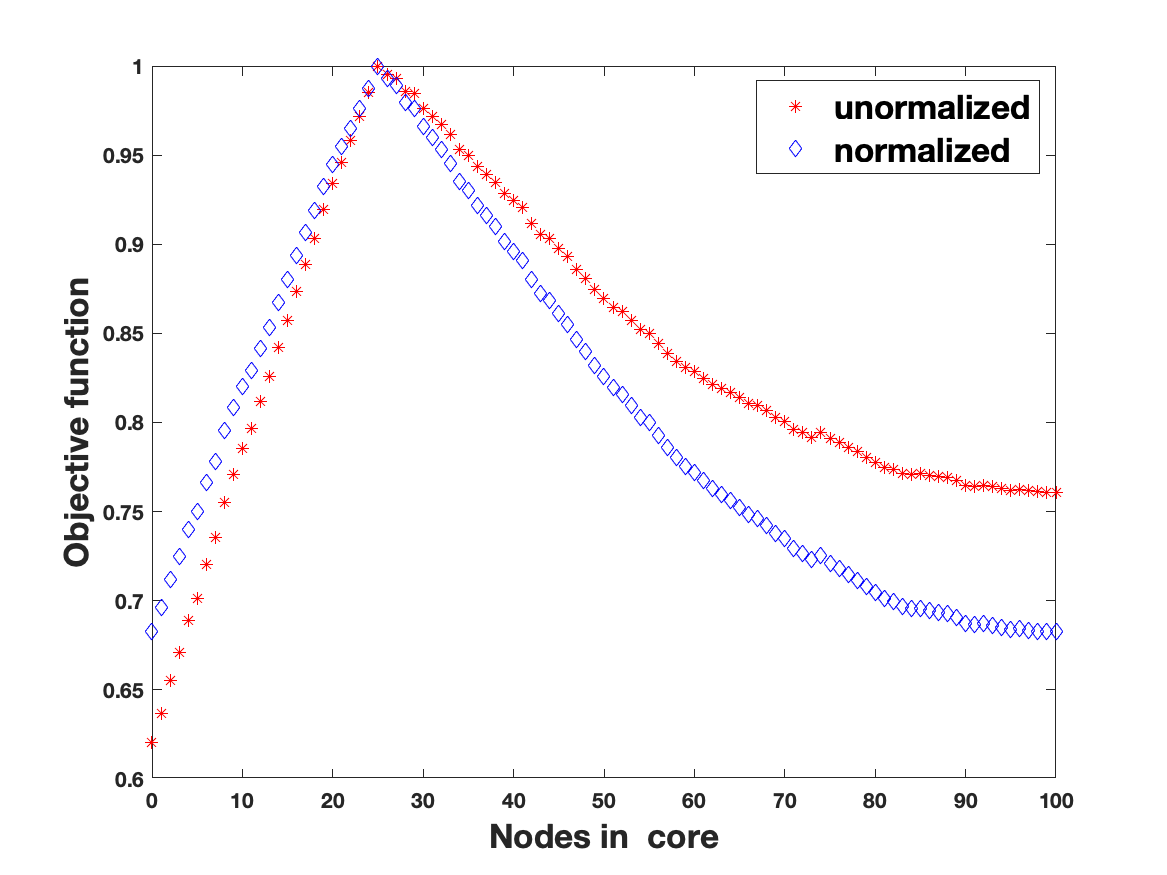}}
\scalebox{0.3}{\includegraphics{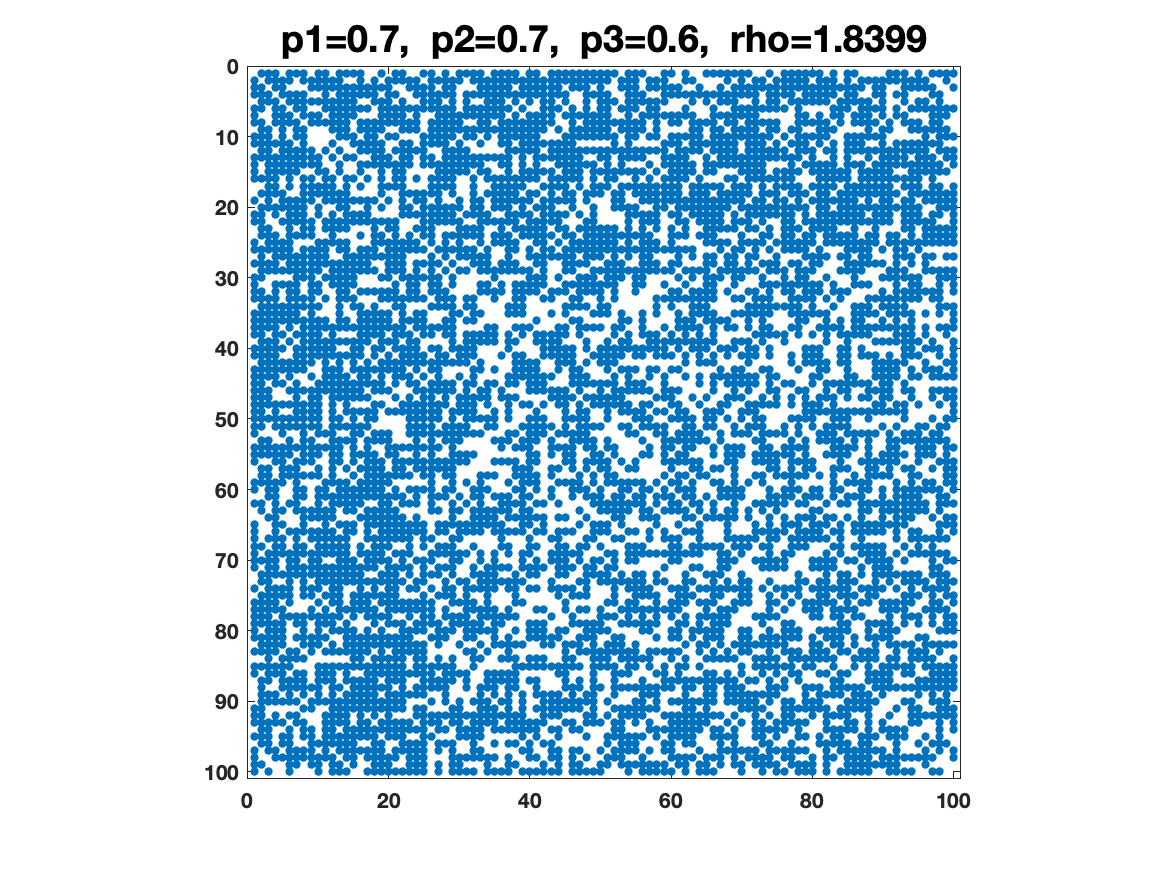}}
\scalebox{0.3}{\includegraphics{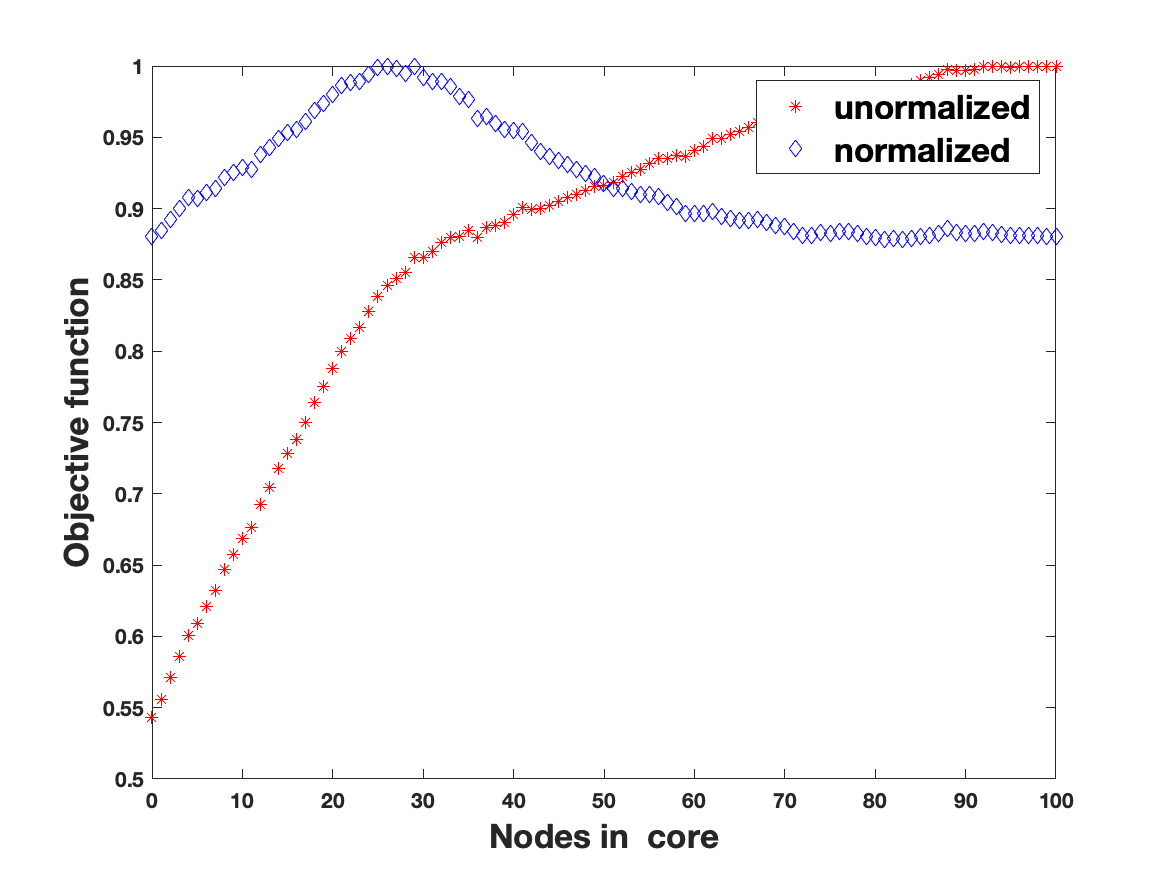}}
\caption{Left: Adjacency matrices with $M = 25$ planted core nodes. 
Right: behaviour of the objective function
(\ref{eq:obj1}) (red stars) and 
(\ref{eq:obj1b}) (blue diamonds) as nodes are added into the core, using the original ordering. Each curve is scaled to have maximum value equal to one.
The value of $\rho$ in (\ref{eq:rhodef}) changes from $ \approx 0.1$,
to $0.4$,
$1.2$ and $1.8$ as we move down the rows.
The normalized objective function (\ref{eq:obj1b}) peaks at the correct core size of 25 in each case. 
}
\label{fig:rhos}
\end{figure}

Figure~\ref{fig:rhos} illustrates 
the difference between 
(\ref{eq:obj1}) and 
(\ref{eq:obj1b}).
Here the networks are samples of a stochastic block model
\cite{cucuringu2016detection,GYW21,rombach2017core,tudisco2019core,ZTM15}.
We will let $\mathrm{SBM}(N,M,p_1,p_2,p_3)$ denote the
stochastic block model with $N$ nodes, a core of size $M$ and 
core-core, core-periphery and 
periphery-periphery probabilities of 
$p_1$, $p_2$ and $p_3$, respectively.
Here, the first $M$ nodes form the core so that
the edge connecting nodes $i$ and $j$, with $i < j$, 
exists with 
independent probability given by 
\begin{description}
\item[core-core:] $p_1$, if $1 \le j \le M$,
\item[core-periphery:] $p_2$, if $1 \le i \le M$ and  $M < j$,
\item[periphery-periphery:] $p_3$ if $M < i$.
\end{description}
In the two upper plots of Figure~\ref{fig:rhos} we sampled from
$\mathrm{SBM}(100,25,0.2,0.2,0.01)$.
We let 
\begin{equation}
 \rho = \frac{N_1}{N_2}
 \label{eq:rhodef}
 \end{equation}
 denote the ratio between the number of edges, $N_1$, and the number of missing edges, $N_2$. 
 The upper left plot in Figure~\ref{fig:rhos} shows the 
 adjacency matrix, with a dot indicating the presence of an edge; here 
$\rho \approx 0.1$.
In the upper right plot, a value of $k$ on the 
horizontal axis represents the partition where, based on the ``correct'' ordering for the SBM, the 
first $k$ nodes are assigned to the core and the 
remaining $N-k$ nodes are assigned to the periphery;
so $x_i = 1$ for $i \le k$ and 
$x_i = 0$ for $k+1 \le i$.  
For each such partition, 
red asterisks 
and blue diamonds show the value of the 
unnormalized objective function 
(\ref{eq:obj1}) and the normalized version 
(\ref{eq:obj1b}). 
Each curve is scaled to have maximum value equal to one. 
In this case, because the overall 
network is sparse, 
the unnormalized measure 
(\ref{eq:obj1})
degrades montonically as we add nodes to the 
core---the scarcity of edges makes it beneficial to 
predict as many missing edges as possible with periphery-periphery
pairs.
So a core of size zero is considered optimal. 
The 
normalized measure 
(\ref{eq:obj1b})
does not suffer from this drawback---here the initial 
addition of nodes into the core gives an increase until 
all 25 ``correct'' nodes are included, after which
the value decreases.

We now alter the probability parameters.
In the second level of Figure~\ref{fig:rhos} the ratio of existing to missing edges is slightly more balanced; we have a sample from 
$\mathrm{SBM}(100,25,0.6,0.5,0.1)$, for which 
$\rho \approx 0.4$.
We see that both measures now give a peak at core size 25, but the normalized version (\ref{eq:obj1b})
gives a more pronounced result.
The third level uses
a sample
from
$\mathrm{SBM}(100,25,0.8,0.8,0.35)$. Here,   
$\rho \approx 1.2$, so the ratio is well balanced.
Both measures are seen to perform effectively.
At the opposite extreme to the first level,
in the fourth level of Figure~\ref{fig:rhos} we have the case of a dense network; here 
we sampled from 
$\mathrm{SBM}(100,25,0.7,0.7,0.6)$, with 
$\rho \approx 1.8$.
We see that the unnormalized measure 
(\ref{eq:obj1}) favours the assignment of all nodes to the core, so that edges are predicted for every pair of nodes.
The normalized version 
(\ref{eq:obj1b})
continues to 
highlight the ``correct'' assignment of the first 25 nodes to the core, even though the 
structure is barely perceptible in the adjacency matrix plot.

In summary, we see that the normalization in 
(\ref{eq:obj1b}) produces a measure that is insensitive to the edge density. This property is 
highly desirable in practice, since networks are typically sparse.
Hence, we will focus on this objective function.

\subsection{Quadratic Form}
\label{subsec:qubo}

Because summing over $1$ and summing over $a_{ij}$ 
in 
(\ref{eq:obj1b}) is not affected by the choice of $x$,  maximizing \eqref{eq:obj1b} is equivalent to maximize
\[
\sum_{i=1}^{N} \sum_{j=1, j \neq i}^{N}
\left(
a_{ij} 
(\frac{1}{N_1} + \frac{1}{N_2} )
-
 \frac{1}{N_2}
 \right)
 \max\{x_i,x_j\}. 
\]
Rescaling by $N_1$ and using
(\ref{eq:rhodef}), we arrive at 
\begin{equation}
\sum_{i=1}^{N} \sum_{j=1, j \neq i}^{N}
\left(
a_{ij} 
(1 + \rho )
-
 \rho
 \right)
 \max\{x_i,x_j\}. 
 \label{eq:obj1c}
\end{equation}
This expression has a direct interpretation: 
for every connected pair of nodes 
$i$ and $j$, where $a_{ij} = 1$,
we gain by $+1$ if the partition correctly predicts 
an edge 
($\max\{x_i,x_j\} = 1$) and by zero otherwise.
Similarly, 
for every disconnected pair of nodes 
$i$ and $j$, where 
$a_{ij} = 0$,
we lose out by $-\rho$ if the partition incorrectly predicts an edge 
($\max\{x_i,x_j\} = 1$) and by zero otherwise.

Since $x$ has binary components, 
we have 
$
\max\{x_i,x_j\} = x_i + x_j - x_i x_j = x_i^2 + x_j^2 - x_i x_j
$,
and hence 
may write the objective function in (\ref{eq:obj1c}) as 
\begin{equation}
\sum_{i=1}^{N} \sum_{j=1, j \neq i}^{N}
\left(
a_{ij} 
(1 + \rho )
-
 \rho
 \right)
 (
 x_i^2 + x_j^2 - x_i x_j).
 \label{eq:newform}
\end{equation}

To find an appropriate QUBO formulation, 
we may expand (\ref{eq:newform})
as 
\[
\sum_{i=1}^{N} 
 x_i^2
\sum_{j=1, j \neq i}^{N}
\left(
a_{ij} 
(1 + \rho )
-
 \rho
 \right)
 +
 \sum_{j=1}^{N} 
 x_j^2
\sum_{i=1, j \neq i}^{N}
\left(
a_{ij} 
(1 + \rho )
-
 \rho
 \right)
-
\sum_{i=1}^{N} 
\sum_{j=1, j \neq i}^{N}
\left(
a_{ij} 
(1 + \rho )
-
 \rho
 \right)
 x_i x_j.
 \]
 Because $A$ is symmetric, 
 the first two terms are equal, and we may rewrite the expression as
 \[
  2 
  \sum_{i=1}^{N} 
 x_i^2
 \left( 
  \mathrm{deg}_i 
(1 + \rho )
- (N-1) \rho \right)
-
\sum_{i=1}^{N} 
\sum_{j=1, j \neq i}^{N}
\left(
a_{ij} 
(1 + \rho )
-
 \rho
 \right)
 x_i x_j.
 \]
 It follows that the maximization of (\ref{eq:newform})
 may be written in QUBO form:
\begin{equation}
    \max_{x_i \in \{0,1\}}
     x^T Q x,
     \quad \text{~where~~~}  
      Q = 2(1+\rho)D - 2(N-1) \rho I
  -A(1+\rho) + \rho E.
     \quad 
    \label{eq:quboform}
\end{equation}
% where 
 %the diagonal elements of $Q \in \RR^{N \times N}$ have the form 
 %\[
 %q_{ii} = 2 ( 
 % \mathrm{deg}_i 
%(1 + \rho )
%- (N-1) \rho )
%\]
%and the off-diagonal elements ($i \neq j$)  have the form
%\[
%q_{ij}
%=
%-a_{ij} (1 + \rho )
%+
% \rho.
 %\]
 %So we may write 
 %\begin{equation}
 % Q = 2(1+\rho)D - 2(N-1) \rho I
 % -A(1+\rho) + \rho E.
 % \label{eq:Qdef}
  % \end{equation}

%[I have checked (\ref{eq:Qdef}) in MATLAB by 
%testing that it is equal to (\ref{eq:obj1b}) up to a shift and scale.]

We note in passing that the coefficient matrix defining a QUBO is not uniquely 
determined; for example, in any QUBO we can force $Q$ to be symmetric, 
upper triangular or lower triangular 
\cite{GKY19}.

\subsection{Modified QUBO Form}

The D-Wave quantum annealer mentioned in section~\ref{sec:dwave}
can handle larger problem dimensions $N$ in 
(\ref{eq:quboform}) if the matrix $Q$ is sparse.
We will assume now that the underlying network represented 
by the adjacency matrix $A$ is sparse, which also implies that 
the ratio $\rho$ in (\ref{eq:rhodef}) is small.
In this case the first three terms in the definition of $Q$ in  
(\ref{eq:quboform}) are sparse, and indeed on  
removing the final term, $\rho E$, the resulting matrix
 \begin{equation}
  \widehat{Q} = 2(1+\rho)D - 2(N-1) \rho I
  -A(1+\rho),
  \label{eq:Qhatdef}
  \end{equation}
has the same sparsity as $A$.

Letting $x_{\mathrm{sum}} := \sum_{i=1}^{N} x_i$,
for any binary-valued $x$ we have 
\[
x^T  E x = x^T \left( \mathbf{1} \mathbf{1}^T - I \right) x = 
 (x^T \mathbf{1} )^2 - x^T x 
 = x_{\mathrm{sum}}^2 - x_{\mathrm{sum}}.
 \]
 Hence, we have 
 \begin{equation}
     x^T Q x - x^T \widehat{Q} x = \rho \, x_{\mathrm{sum}} ( x_{\mathrm{sum}} -1).
     \label{eq:Qdiff}
     \end{equation}
     At an optimal value of $x$; that is, a binary vector 
     maximizing (\ref{eq:quboform}), the quantity
    $x_{\mathrm{sum}}$
     represents the number of nodes assigned to the core.
      For a sparse network we expect the core size to be 
      small compared with $N$.
   Hence, the difference in (\ref{eq:Qdiff}) should be
    small relative to $x^T Q x$.  So the 
    original QUBO
     (\ref{eq:quboform}), which has a full matrix $Q$, should 
     be well approximated by the sparse QUBO
  \begin{equation}
    \max_{x_i \in \{0,1\}}
     x^T \widehat{Q} x.
    \label{eq:quboform2}
\end{equation}   
Based on this motivation,
for large sparse networks 
where the original 
QUBO
(\ref{eq:quboform})
cannot be treated by D-Wave, we will use the 
nearby sparse QUBO
(\ref{eq:quboform2}).
However, for consistency we will judge the quality of the solution 
 in terms of the original quadratic form $x^T Q x$ as in  (\ref{eq:quboform}).

\section{Quantum Annealing}
\label{sec:dwave}

Quantum annealers may only be applied to problems in  
QUBO form (or an equivalent Ising form).
Although this restricts their practical usage, we note that
many tasks arising in graph theory, scheduling and 
theoretical computer science
may be expressed as QUBOs;  see  
\cite{CTV21,CDHR19} for recent examples and 
\cite{GKY19,Lucas14} for comprehensive reviews.

 The essence of quantum annealing is to move  
 adiabatically from a ``simple'' Hamiltonian 
 to a Hamiltonian that encodes the problem of interest.
 This annealing process makes use of quantum phenomena, including superposition and tunneling, to explore the solution landscape.
 %If the quantum system remains in the ground state 
 %throughout, then the final
 %measurement will return a solution of the problem.

 As discussed in \cite{McG20},
because quantum annealing takes place in a physical system 
 it is subject to ambient noise and liable to suffer further imprecision resulting from the analog controls. 
 For these reasons it is difficult to make general 
 statements about either the theoretical computational complexity 
 or the practical performance of a quantum annealer.
 However, there are indications \cite{CL21} that quantum annealing, 
 and quantum computing in general \cite{A19},
 have the potential to make a larger range of 
 problems computationally feasible.
 Our approach in this work is to focus 
 on the quality 
 of solution 
 provided 
 by the quantum annealer 
 and to compare this with the results obtained by existing heuristic 
 approaches on a classical machine.

Our quantum annealing experiments are conducted on the  
Advantage 4.1 system from 
D-Wave  
\cite{MF21}, which is commercially available via remote access.
%At the end of each computation, a single solution is chosen at random from a set of good solutions, and returned. 
The output is probabilistic, and hence it is common to
request multiple samples for comparison.
In our computations, we found that 100 samples 
was sufficient to provide consistent results. 
By default we will report on the best sample obtained. 

\begin{figure}[htbp]
\centering
\scalebox{0.4}{
\includegraphics{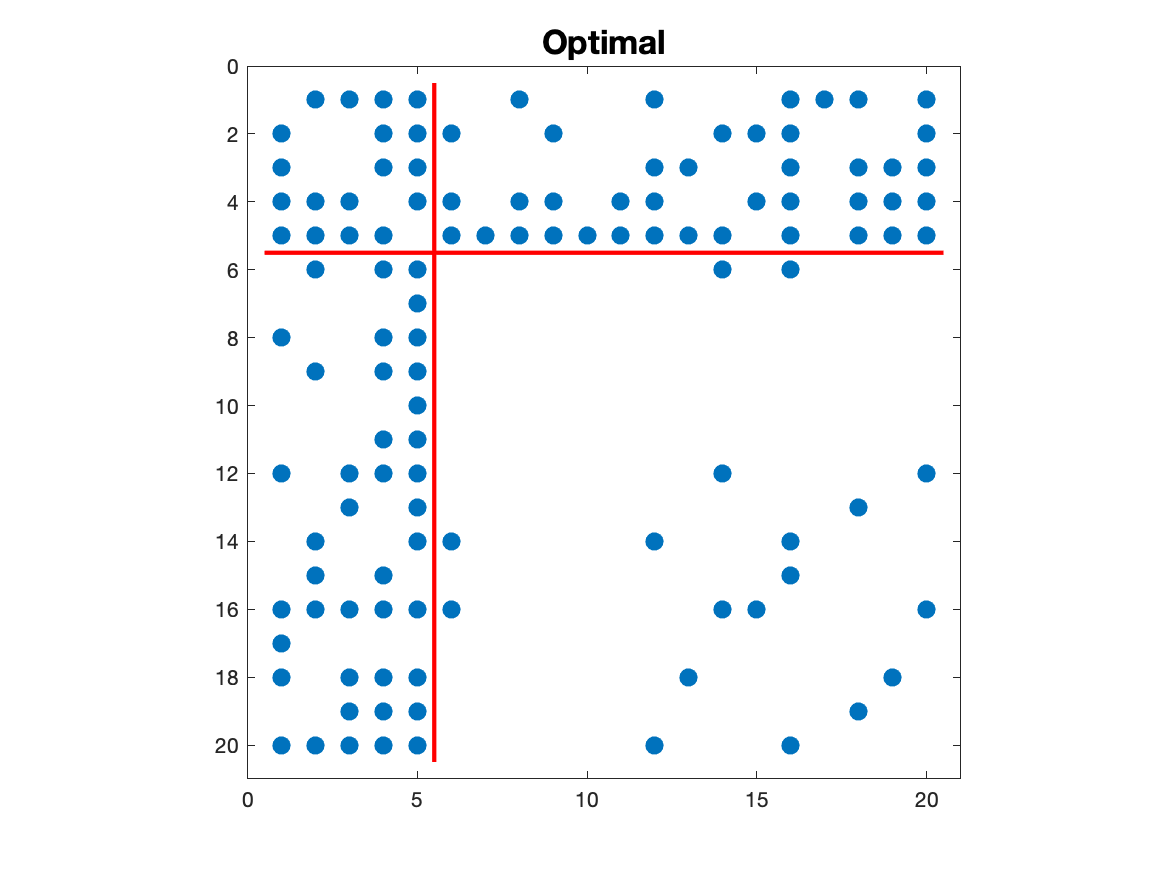}}
\scalebox{0.4}{
\includegraphics{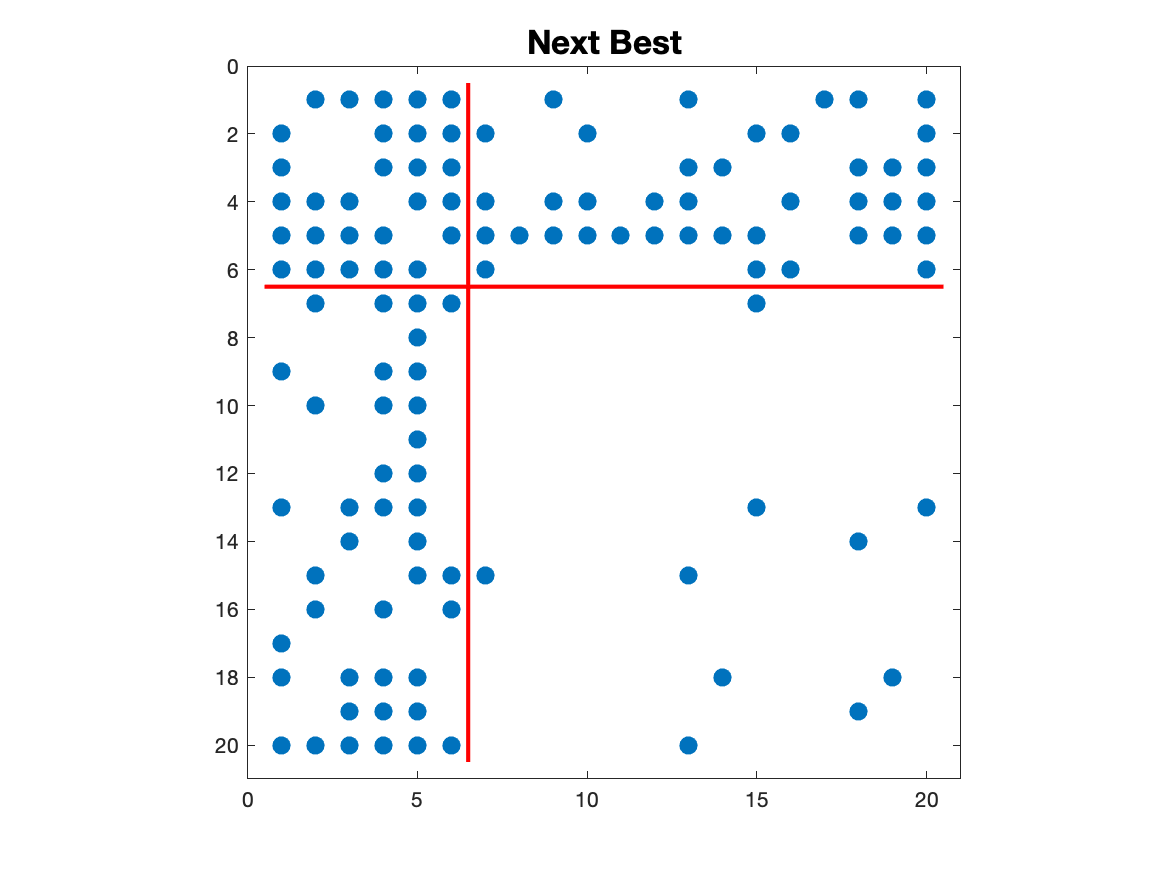}}
\caption{Left: Adjacency matrix with core-periphery partition from 
\cite{borgatti2000models}.
Right: second best  core-periphery partition from the quantum annealer added an extra node to the core.}
\label{fig:journal}
\end{figure}

For illustration, 
on the left in Figure~\ref{fig:journal} we show the adjacency matrix for a network 
used in \cite[Table~4]{borgatti2000models}
concerning co-citations among social work journals. 
Here, the horizontal and vertical lines 
illustrate the partition proposed
in \cite{borgatti2000models} using a genetic algorithm to maximize the correlation between the data and an
ideal pattern matrix. We see that five nodes have
been placed in the core. 
In this case, the quantum annealer applied to the corresponding
QUBO (\ref{eq:quboform}) 
produced the same 
core-periphery
partitioning, thereby supporting the empirical result in 
\cite{borgatti2000models}.
For this partition, with $Q$ defined in (\ref{eq:quboform}),
the objective function has the value 
$x^T Q x \approx 68.5$. 
For information, on the right in
Figure~\ref{fig:journal}
we also report the 
second-best
partition returned by the quantum annealer; here an extra node 
(the original node $16$) has been placed in the core and the objective function value is 
$x^T Q x \approx 67.5$.

\begin{figure}[htbp]
\centering
\scalebox{0.4}{
\includegraphics{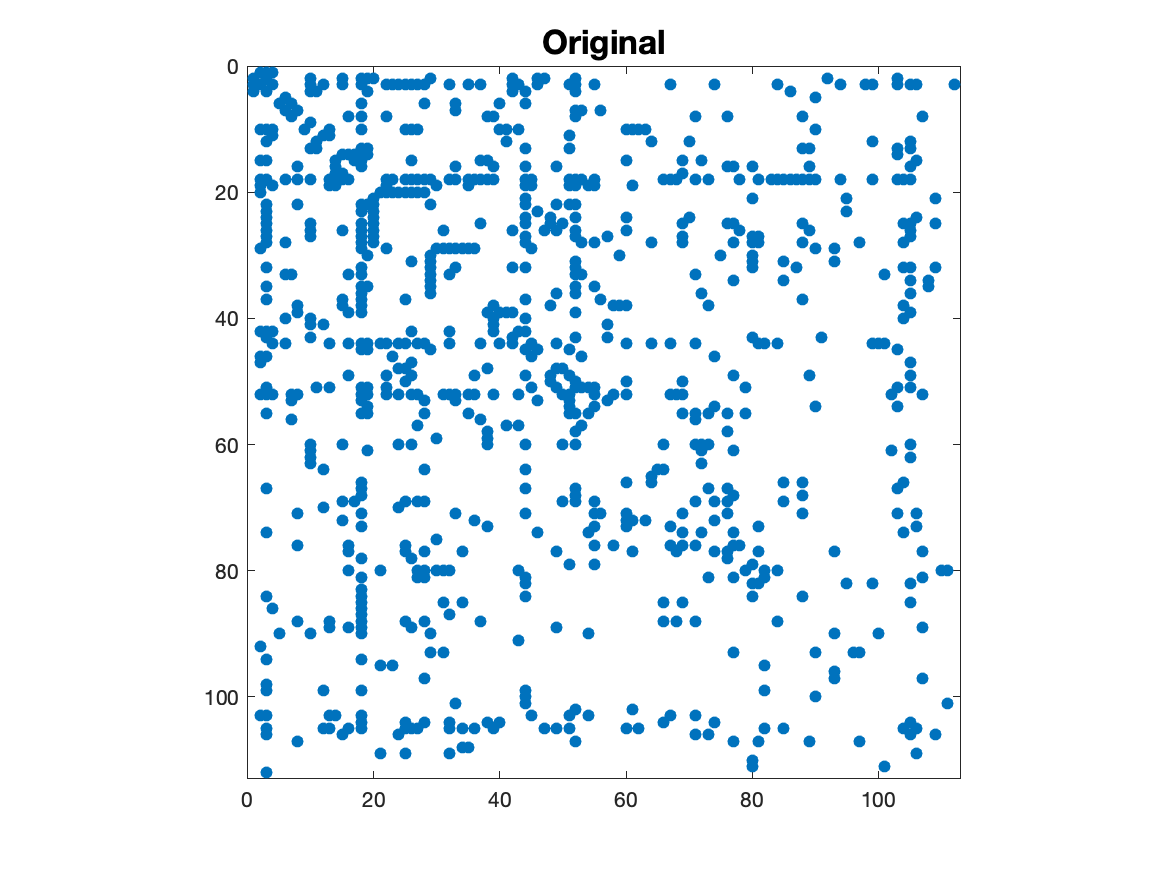}}
\scalebox{0.4}{
\includegraphics{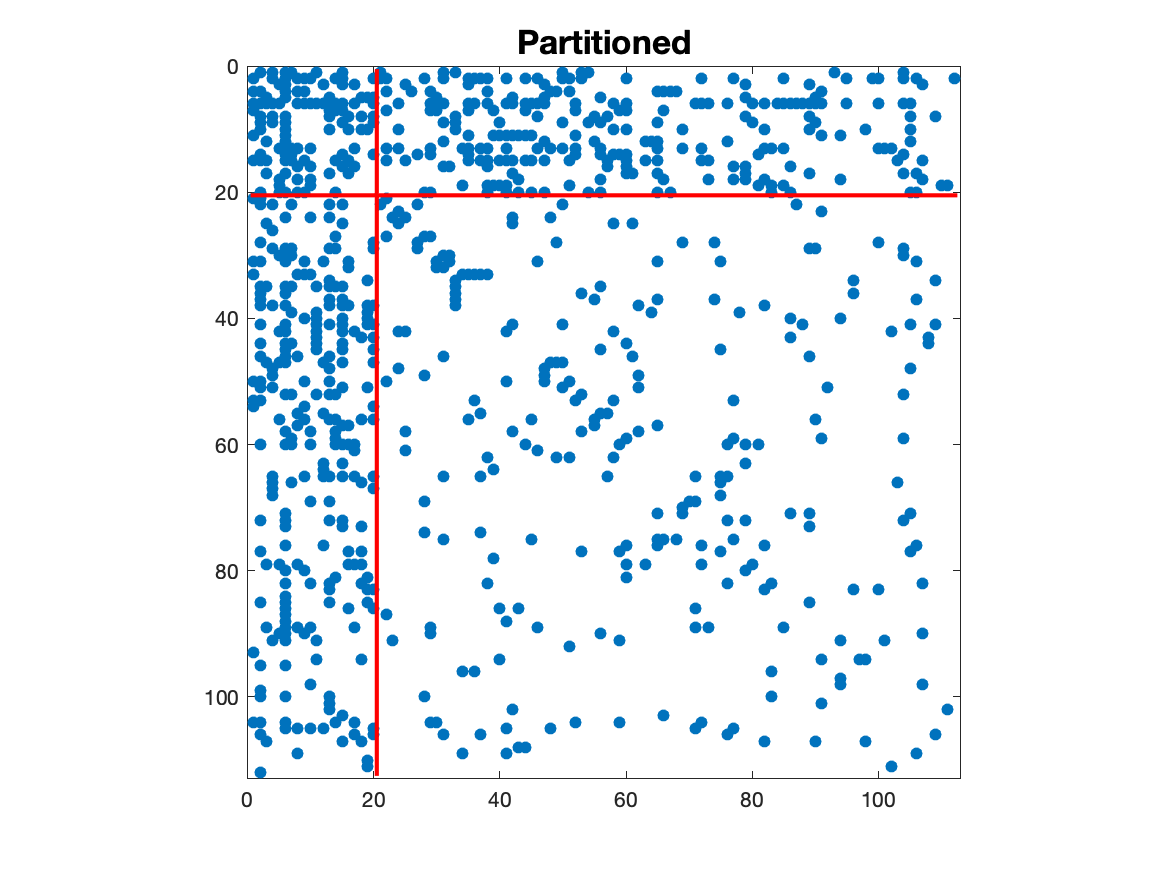}}
\caption{Left: Adjacency matrix from \cite{Newman06} with original node ordering.
Right: best core-periphery partition from the quantum annealer.}
\label{fig:adjnoun}
\end{figure}

As a further illustration, 
in 
Figure~\ref{fig:adjnoun} the nodes represent 
the 60 most commonly occurring adjectives and the 
60 most commonly occurring 
nouns
in the novel ``David Copperfield'' by Charles Dickens, 
and edges connect pairs of words that occur in adjacent positions in the text. 
(Eight nodes are disconnected from the rest of the network, and hence are ignored.) 
This network data, with 425 edges,  comes from   
\cite{Newman06}.
The picture on the left shows the 
adjacency matrix in the original ordering, and on the 
right we give the best core-periphery partition 
found by the quantum annealer.
Further results for this network appear in Table~\ref{tab:Qreal} 
of section~\ref{sec:comp}.

\section{Method Comparison}
\label{sec:comp}

In this section we give quantitative results 
based on the new 
objective function
(\ref{eq:obj1c}), via the quadratic form in 
(\ref{eq:quboform}), 
as a means to compare various approaches to
core-periphery partitioning.

\subsection{Synthetic Data}
\label{subsec:synth}

\begin{figure}[htbp]
\centering
\scalebox{0.4}{
\includegraphics{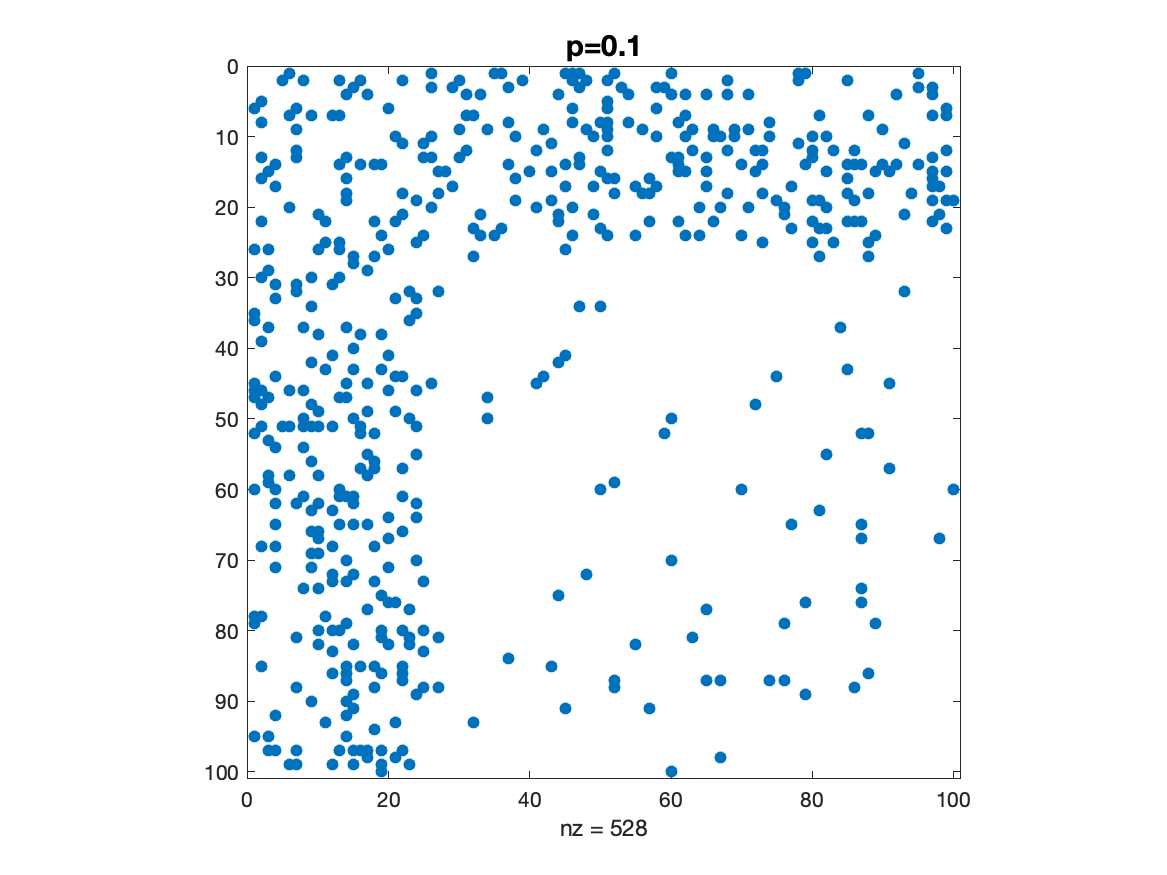}}
\scalebox{0.4}{
\includegraphics{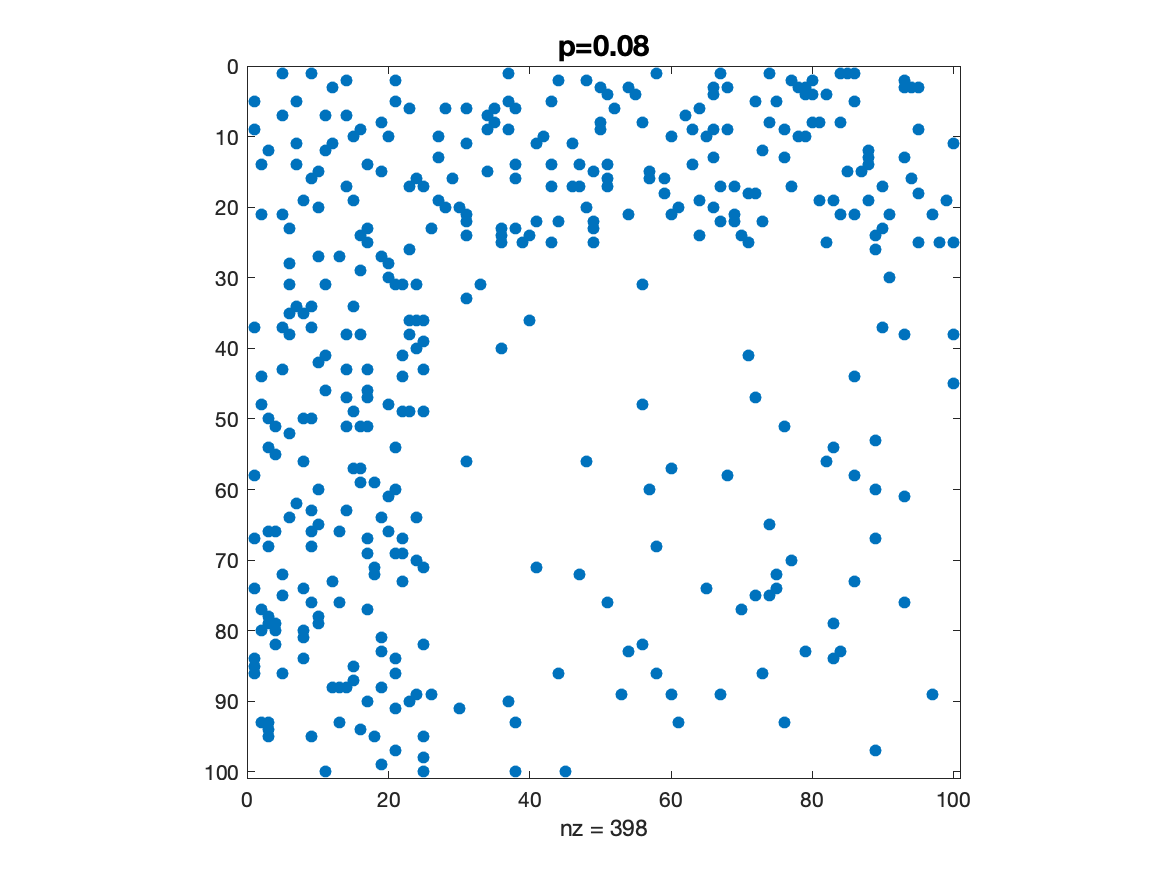}}
\scalebox{0.4}{
\includegraphics{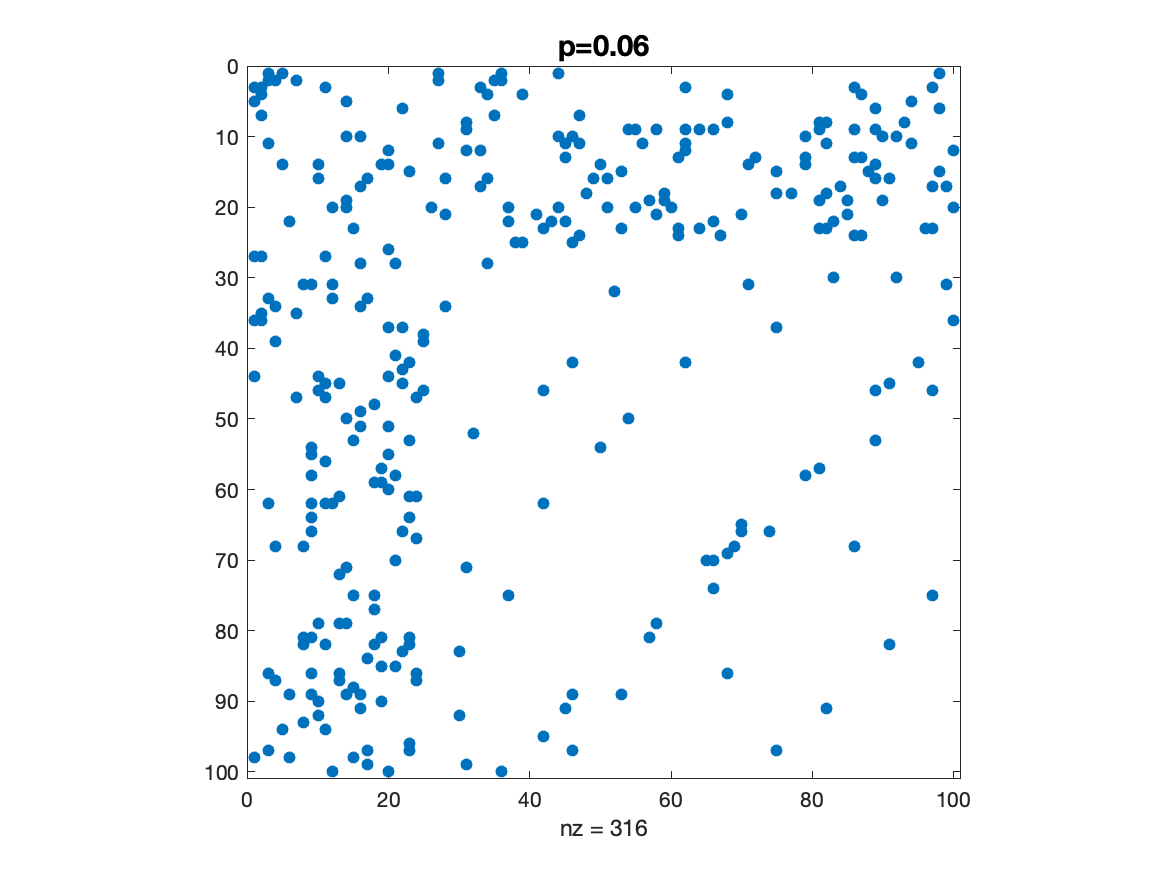}}
\scalebox{0.4}{
\includegraphics{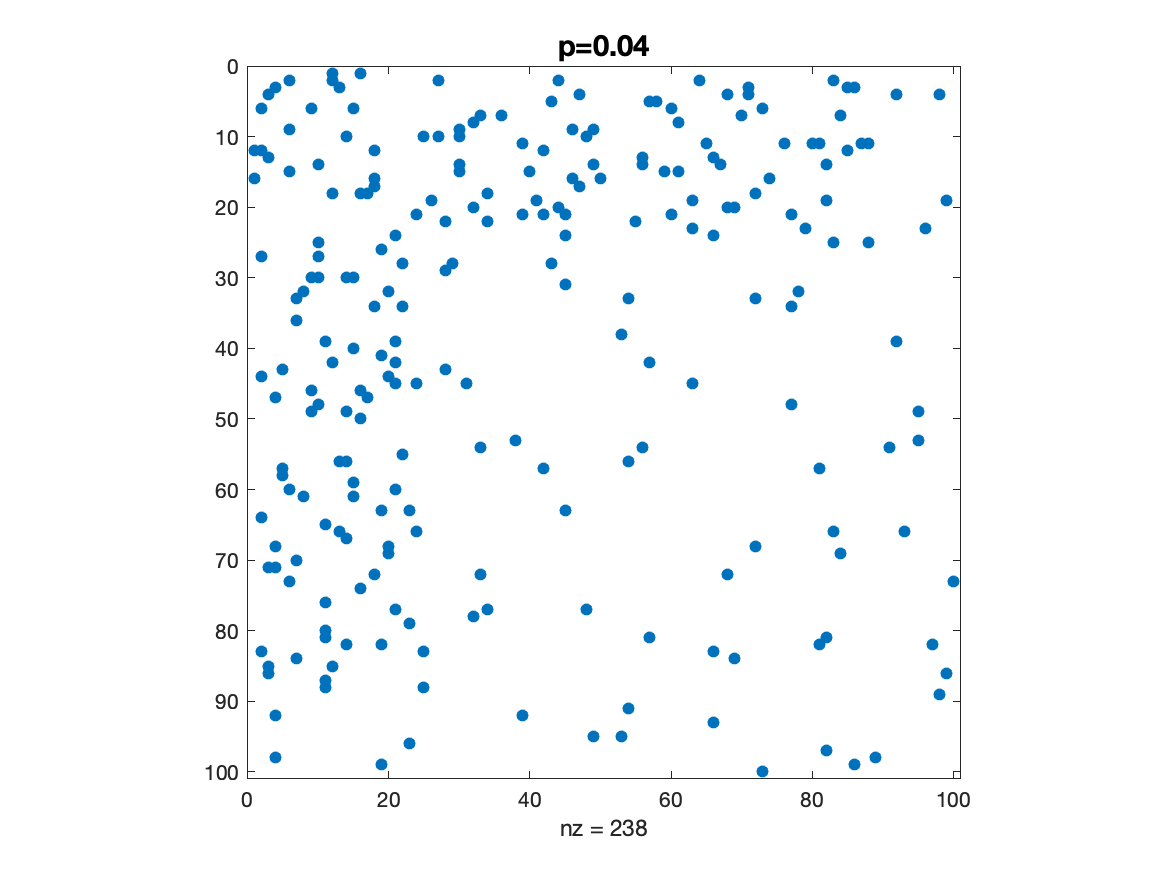}}
\caption{Small synthetic data: samples from
$\mathrm{SBM}(100,25,p,p,0.01)$
    for
     $p = 0.1$, $p = 0.08$, $p = 0.06$ and $p = 0.04$.
     }
\label{fig:small}
\end{figure}

We begin with two tests on stochastic block models 
that have some level of  planted core-periphery structure.
In Figure~\ref{fig:small} we show samples
from 
 $\mathrm{SBM}(100,25,p_1,p_2,p_3)$
    with $p_1 = p_2 = p$ and $p_3 = 0.01$, for 
     $p = 0.1$, $p = 0.08$, $p = 0.06$ and $p = 0.04$.
Table~\ref{tab:Q1} records the results.
Here, each partitioning method produces a binary vector,
$x \in \RR^{100}$, and we show the 
corresponding 
value of $x^T Q x$ for $Q$ in (\ref{eq:quboform}).
In parentheses we show the associated core size,
$x_{\mathrm{sum}}$. 
For reach network, the two 
largest values of $x^T Q x$ are highlighted, with 
the largest value shown in bold.
The first row, marked ``Original'',
shows the value of $x^T Q x$ arising when $x$ is taken to be
the ``correct'' core set arising from the model; that is,
$x_i = 1$ for $1 \le i \le 25$ and $x_i = 0$ otherwise.
We emphasize that due to the stochasticity this partitioning
is not guaranteed to be optimal for a particular SBM sample; indeed,
we see from the table that better choices exist in each case.
The second and third rows, marked ``$Q$'' and 
 ``$\widehat{Q}$'',
 show results for the 
 D-Wave solution to the QUBO
 (\ref{eq:quboform})
 and 
 (\ref{eq:quboform2}), respectively. Rows 
 four to nine correspond to techniques originally designed to output a vector of nonnegative values to be regarded as a measure of coreness. From these vectors, we find a binary partitioning vector $x$ by optimally thresholding  the  coreness vector: we assign value $x_i=1$ to the top $k$ nodes in terms of coreness, and $x_i=0$ to the remaining nodes, and then select the binary $x$ which attains the largest value of $x^T Q x$ over $0 \le k \le N$. The corresponding $k$ is then taken to be the predicted core size.
 The fourth row, ``Degree,'' uses the degree vector 
$\mathrm{deg}_i$ as a measure of coreness. 
Similarly, the fifth, sixth and seventh rows, 
``EigA,'' 
``EigQ'' 
and ``NonlinPM''
use optimal partitions 
based on the coreness 
measures given by the Perron-Frobenius 
eigenvector of $A$,
the dominant 
eigenvector of $Q$,
and the corresponding nonlinear eigenvector from
\cite{tudisco2019core} (with parameter values $\alpha = 10$ and
$p = 2\alpha$
taken from that work).
We note that the use of the degree vector and 
the Perron-Frobenius vector of $A$ 
was suggested in \cite{borgatti2000models}
and has subsequently been studied by several authors; see for example, 
\cite{rombach2017core,tudisco2019core,ZTM15}.
The ``EigQ'' method is motivated by the idea of using 
an eigenvector that solves a relaxed
version of the QUBO (\ref{eq:quboform}), where
$x \in \RR^{N}$ is constrained to have $\| x \|_2 = 1$. The eighth row ``$h$-index'' uses the $k$-core decomposition coreness score \cite{kitsak2010identification}, computed as the limit of the $h$-index operator sequence \cite{lu2016h}, while the ninth row ``GenBE'' corresponds to the generalization of the original Borgatti and Everett core measure \cite{borgatti2000models} proposed in \cite{rombach2017core}, where the quadratic form $x^T A x$ is approximately maximimzed over a set of not-necessarily-binary core-periphery transition vectors $x$. 

\renewcommand{\arraystretch}{1.2}
\begin{table}[t]
    \centering
    \begin{tabular}{lcccc}
    \toprule
    & $p = 0.1$ &  $p = 0.08$ &  $p = 0.06$ &  $p = 0.04$ 
 \\ \midrule
    Original       &  245.1 (25)   & 165.8 (25) & 131.3 (25) & 83.4 (25)  \\ 
    $Q$            & \first{255.7} (23) & \first{174.5} (28) & 
    \first{183.4} (24) & \first{101.3} (28) \\
    $\widehat{Q}$  & \second{252.1} (21) & \second{174.2} (24) & \second{183.1} (22) & 98.8  (22)      \\
    Degree  &  241.0 (26) & 164.1 (28) & 136.3 (24) & 89.2 (24)     \\
    EigA  &  177.6 (23) & 89.7 (20) & 88.6 (27) & 49.5 (27)      \\
    EigQ  &  119.0 (21) & 86.6 (321) & 59.0 (25) & 70.0 (33)      \\
    NonlinPM  & \first{255.7} (23)  & 172.4 (28) & 136.8 (26) & \second{100.5} (30)  \\
    $h$-index & 235.2 (21) & 165.8 (25) & 118.6 (19) &  81.0 (24) \\
    GenBE & 150.3 (31) &  91.4 (21) &  90.0 (25) &  56.0 (23) \\
    \bottomrule
\end{tabular}
%     \begin{tabular}{lllll}
%     \toprule
%     & $p = 0.1$ &  $p = 0.08$ &  $p = 0.06$ &  $p = 0.04$ 
%  \\ \midrule
%     Original       &  245.1 (25)   & 165.8 (25) & 131.3 (25) & 83.4 (25)  \\ 
%     $Q$            & \textbf{255.7} (23) & \textbf{174.5} (28) & 
%     \textbf{183.4} (24) & \textbf{101.3} (28) \\
%     $\widehat{Q}$  & 252.1 (21) & 174.2 (24) & 183.1 (22) & 98.8  (22)      \\
%     Degree  &  241.0 (26) & 164.1 (28) & 136.3 (24) & 89.2 (24)     \\
%     EigA  &  177.6 (23) & 89.7 (20) & 88.6 (27) & 49.5 (27)      \\
%     EigQ  &  119.0 (21) & 86.6 (321) & 59.0 (25) & 70.0 (33)      \\
%     NonlinPM  & \textbf{255.7} (23)  & 172.4 (28) & 136.8 (26) & 100.5 (30)  \\
%     $h$-index & 235.2 (21) & 165.8 (25) & 118.6 (19) &  81.0 (24) \\
%     GenBE & 150.3 (31) &  91.4 (21) &  90.0 (25) &  56.0 (23) \\
%     \bottomrule
% \end{tabular}
    \caption{
    Values of the objective function value $x^T Q x$
    from (\ref{eq:quboform}) for the 
    $\mathrm{SBM}(100,25,p,p,0.01)$ samples  
    shown in Figure~\ref{fig:small}, with core size in parentheses.
Original: assigning the first $M = 25$ nodes to the core;
$Q$: 
quantum annealing on 
(\ref{eq:quboform});
$\widehat{Q}$: 
quantum annealing on 
(\ref{eq:quboform2});
Degree: the nodal degrees;
EigA: the eigenvector associated with 
dominate eigenvalue of the adjacency matrix $A$;
EigQ: the eigenvector associated with 
dominate eigenvalue of the QUBO matrix $Q$;
NonlinPM: the power method from
\cite{tudisco2019core} to compute a nonlinear eigenvector;
$h$-index: the $k$-core decomposition coreness score from \cite{kitsak2010identification};
GenBE: the method from \cite{rombach2017core}. 
For reach network, the two 
largest values of $x^T Q x$ are highlighted, with 
the largest value shown in bold.
}
    \label{tab:Q1}
\end{table}

We see in Table~\ref{tab:Q1} that, on these tests, 
the largest or joint-largest value of $x^T Q x$ is achieved by 
applying the quantum annealing algorithm directly to the 
QUBO (\ref{eq:quboform}).
We also note that 
 $Q$, $\widehat{Q}$ and NonlinPM improve on 
the $x^T Q x$ value provided by the planted ``ground truth''
from the original model.
The two standard eigenvalue approaches are consistently the poorest in this measure.

\begin{figure}[htbp]
\centering
\scalebox{0.3}{
\includegraphics{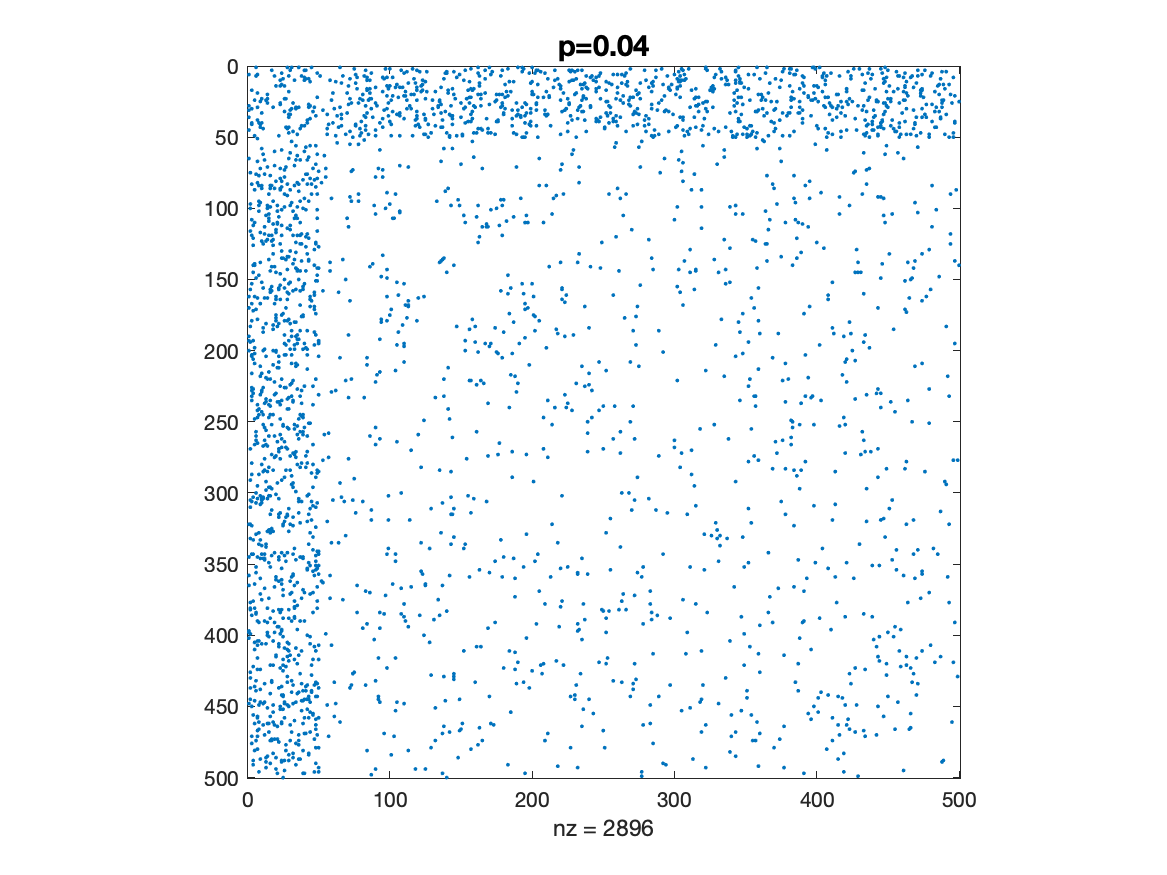}}
\scalebox{0.3}{
\includegraphics{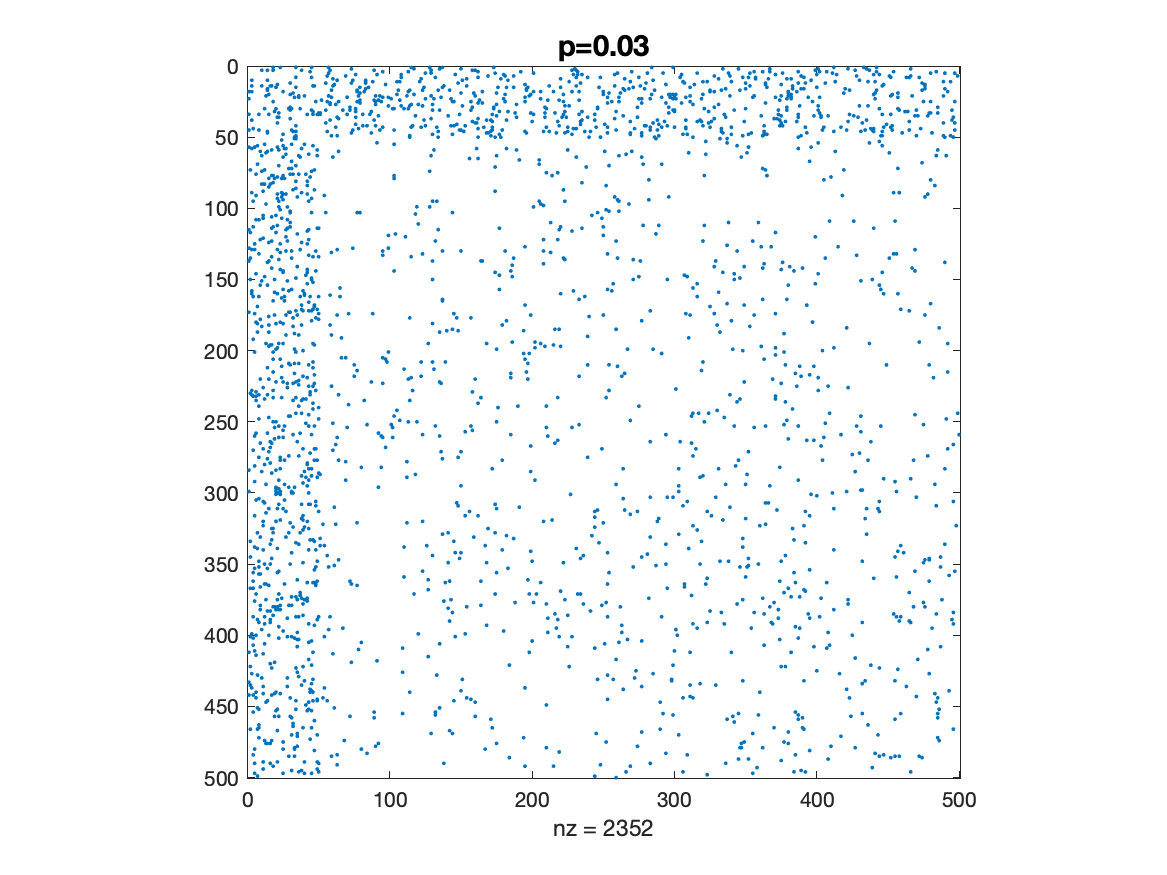}}
\scalebox{0.3}{
\includegraphics{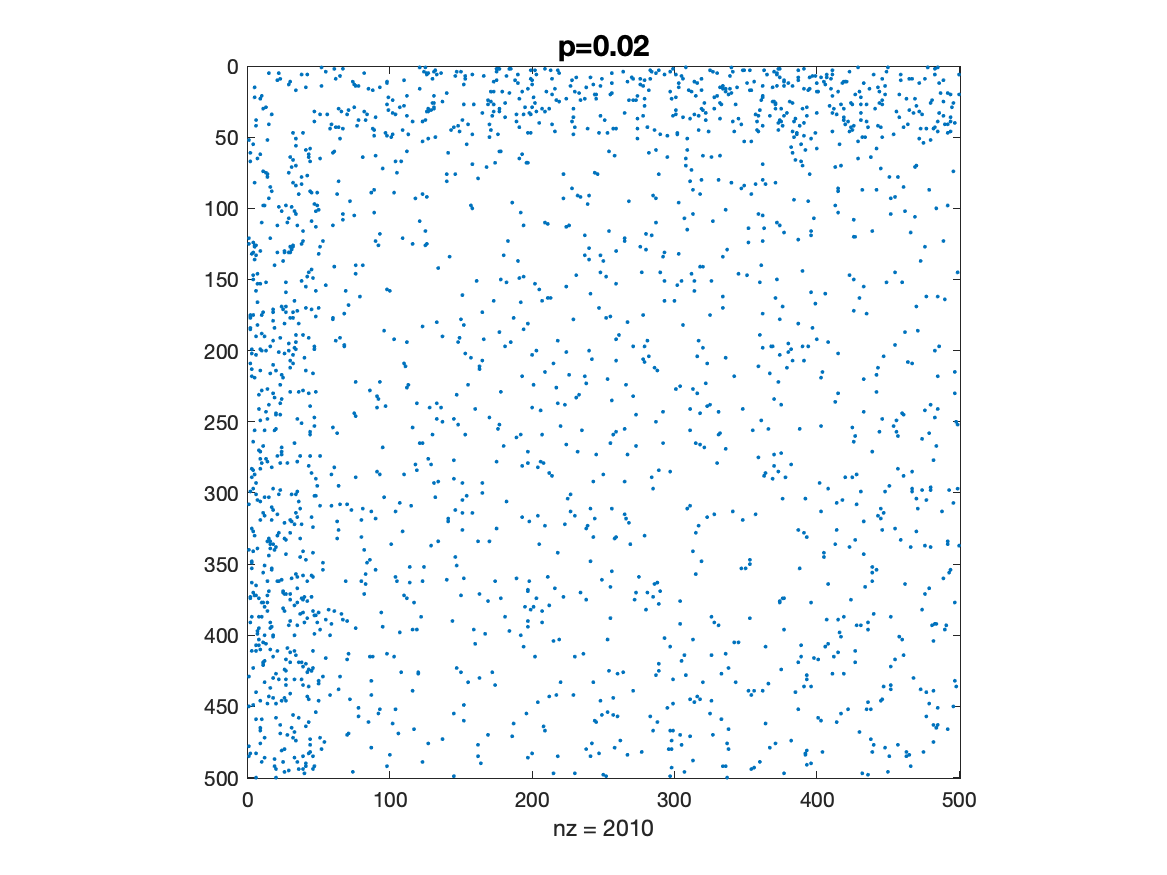}}
\scalebox{0.3}{
\includegraphics{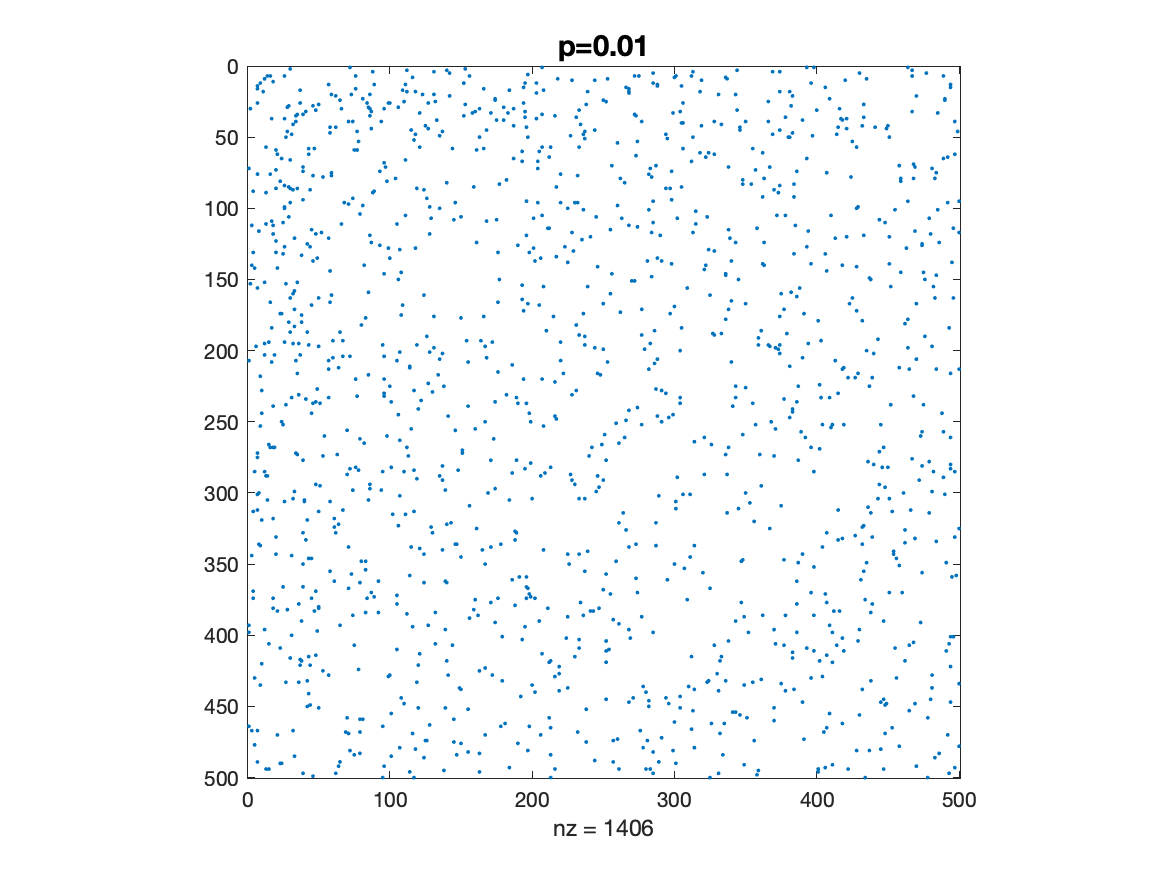}}
\caption{Larger synthetic data: samples from
$\mathrm{SBM}(500,50,p,p,0.005)$
    for
     $p = 0.04$, $p = 0.03$, $p = 0.02$ and $p = 0.01$.
     }
\label{fig:large}
\end{figure}

In Figure~\ref{fig:large} we show larger networks: 
these are samples from 
$\mathrm{SBM}(500,50,p,p,0.005)$
    with 
     $p = 0.04$, $p = 0.03$, $p = 0.02$ and $p = 0.01$.
     Here, the full 
     QUBO (\ref{eq:quboform}) was too large for the quantum annealer. 
     In Table~\ref{tab:Q2} we show results for the remaining 
     partioning methods.
     We see that quantum annealing with $\widehat{Q}$ 
     gives the best result on three of the four cases, 
     with NonlinPM also performing well.

\begin{table}[t]
    \centering
    \begin{tabular}{lllll}
    \toprule
    & $p = 0.04$ &  $p = 0.03$ &  $p = 0.02$ &  $p = 0.01$ 
 \\ \midrule
    Original       &  1334.7 (50)   & 939.6 (50) & 600.6 (50) & 155.5 (50)  \\ 
    $\widehat{Q}$  & \first{1365.5} (62) & \second{996.2} (72) & \first{730.0} (88) & \first{472.3} (120)      \\
    Degree  &  1350.7 (56) & 970.4 (68) &  670.3 (102) & 368.7 (94)     \\
    EigA  &  1213.3 (66) & 726.9 (87) & 431.2 (75) & 189.9 (84)      \\
    EigQ  &  768.8 (111) & 428.9 (111) & 219.8 (196) & 362.3 (198)      \\
    NonlinPM  &  \second{1363.1} (72) & \first{1002.2} (90) & \second{729.2} (101) & \second{465.7} (157)     \\
    $h$-index & 1334.7 (50) & 920.4 (48) & 605.4 (51) & 206.0 (116)\\ 
    GenBE & 1320.1 (57) & 915.4 (76) & 554.7 (99) & 290.0 (151) \\
    \bottomrule
\end{tabular}
%     \begin{tabular}{lllll}
%     \toprule
%     & $p = 0.04$ &  $p = 0.03$ &  $p = 0.02$ &  $p = 0.01$ 
%  \\ \midrule
%     Original       &  1334.7 (50)   & 939.6 (50) & 600.6 (50) & 155.5 (50)  \\ 
%     $\widehat{Q}$  & \textbf{1365.5} (62) & 996.2 (72) & \textbf{730.0} (88) & \textbf{472.3} (120)      \\
%     Degree  &  1350.7 (56) & 970.4 (68) &  670.3 (102) & 368.7 (94)     \\
%     EigA  &  1213.3 (66) & 726.9 (87) & 431.2 (75) & 189.9 (84)      \\
%     EigQ  &  768.8 (111) & 428.9 (111) & 219.8 (196) & 362.3 (198)      \\
%     NonlinPM  &  1363.1 (72) & \textbf{1002.2} (90) & 729.2 (101) & 465.7 (157)     \\
%     $h$-index & 1334.7 (50) & 920.4 (48) & 605.4 (51) & 206.0 (116)\\ 
%     GenBE & 1320.1 (57) & 915.4 (76) & 554.7 (99) & 290.0 (151) \\
%     \bottomrule
% \end{tabular}
    \caption{As in 
    Table~\ref{tab:Q1} for the $\mathrm{SBM}(500,50,p,p,0.005)$
    samples shown in Figure~\ref{fig:large}.
     Here the networks are too large for quantum annealing on the 
     full QUBO (\ref{eq:quboform}).
}
    \label{tab:Q2}
\end{table}

\subsection{Real Data}
\label{subsec:real}

Table~\ref{tab:Qreal} shows results 
for the following real networks:
\begin{description}
\item[USAir97] is from 
\cite{DH11}, with weights binarized.
The $N= 332$ nodes represent airports in USA. The 2126 undirected edges
indicate whether at least one 
scheduled USAir flight took place between the two airports in 1997.
\item[Celegans] has 277 nodes and 2105 edges that represent
neurons and synapses in the worm Caenorhabditis elegans.
The data is from \verb5https://www.cs.cornell.edu/~arb/data/5 based on 
\cite{GD03}.
\item[Jazz] from \cite{CMK04} is a network of 198 jazz bands (nodes) that performed between 1912 and 1940,  and 
2742 corresponding edges (musicians).
\item[Adjnoun] was described in section~\ref{sec:dwave}.
\item[Football] from \cite{girvan2002community}, is a network of American football games between 115 Division IA colleges during the
fall 2000 regular season. Here, nodes represent teams and the 613 edges
represent fixtures.
\item[Journals] from \cite{pajek_data},
has 5972 edges representing shared interests among 124 magazines and journals, which form the nodes, based on a sample of  $\sim 100,000$  residents of Ljubljana (Slovenia) in survey conducted in 1999 and 2000.
\end{description}

\begin{table}[t]
    \centering
    \begin{tabular}{lllllll}
    \toprule
    & USAir97 & Celegans & Jazz & Adjnoun & Football & Journals 
 \\ \midrule
  $Q$  & x  & x &  x & \first{352.5} (20) &  \first{93.8} (51)  &   \first{7360.5} (54)  \\
    $\widehat{Q}$  & \first{2703.7} (38) & \first{1620.3} (42)  & 1743.1 (39) & \second{351.8} (18) &  33.6 (12) 
      & 4091.6 (23)  \\
    Degree  &  2677.1 (35) & 1601.7 (49) &  \second{1772.6} (51) &  345.1 (23)  & 27.9 (38)  & \first{7360.5} (54)   \\
    EigA  & 2598.3 (37) & 1243.8 (21)  & 1400.6 (37)  & 314.5 (19) &  6.4 (4)  
    & 7355.2 (52)\\
    EigQ  & 310.7 (5) & 1414.2 (40)  & 541.6 (16)  & 158.2 (13) & 77.5 (54)  
    & 3741.1 (23)\\
    NonlinPM  &  2699.5 (38) & 501.4 (28)  & \first{1792.4} (47) &  345.3 (22)   & \second{76.4} (38)  & \second{7358.8} (53)  \\
    $h$-index & 2534.9 (42) & 258.8 (18) & 884.4 (60) & 242.7 (32) &  14.1 (89) & 4156.1 (39) \\
    GenBE & 2626.1 (34) & 815.8 (35) & 1364.3 (30) & 298.4 (21) & 3.9 (2) & 7349.6 (53) \\
    \bottomrule
    \end{tabular}
%     \begin{tabular}{lllllll}
%     \toprule
%     & USAir97 & Celegans & Jazz & Adjnoun & Football & Journals 
%  \\ \midrule
%   $Q$  & x  & x &  x & \textbf{352.5} (20) &  \textbf{93.8} (51)  &   \textbf{7360.5} (54)  \\
%     $\widehat{Q}$  & \textbf{2703.7} (38) & \textbf{1620.3} (42)  & 1743.1 (39) & 351.8 (18) &  33.6 (12) 
%       & 4091.6 (23)  \\
%     Degree  &  2677.1 (35) & 1601.7 (49) &  1772.6 (51) &  345.1 (23)  & 27.9 (38)  & \textbf{7360.5} (54)   \\
%     EigA  & 2598.3 (37) & 1243.8 (21)  & 1400.6 (37)  & 314.5 (19) &  6.4 (4)  
%     & 7355.2 (52)\\
%     EigQ  & 310.7 (5) & 1414.2 (40)  & 541.6 (16)  & 158.2 (13) & 77.5 (54)  
%     & 3741.1 (23)\\
%     NonlinPM  &  2699.5 (38) & 501.4 (28)  & \textbf{1792.4} (47) &  345.3 (22)   & 76.4 (38)  & 7358.8 (53)  \\
%     $h$-index & 2534.9 (42) & 258.8 (18) & 884.4 (60) & 242.7 (32) &  14.1 (89) & 4156.1 (39) \\
%     GenBE & 2626.1 (34) & 815.8 (35) & 1364.3 (30) & 298.4 (21) & 3.9 (2) & 7349.6 (53) \\
%     \bottomrule
%     \end{tabular}
    \caption{As in 
    Table~\ref{tab:Q1} for the 
    real networks described in the text.
     The symbol ``x'' denotes that the problem (\ref{eq:quboform}) 
      was too large for the quantum annealer.}
    \label{tab:Qreal}
\end{table}

We see that on the three smaller networks in 
Table~\ref{tab:Qreal}, 
quantum annealing with $Q$ produced the best or joint-best
results.
The x symbol indicates that the first three problems 
produced a QUBO (\ref{eq:quboform}) that was too large for 
the quantum annealer.
Two of the top results on the three larger networks are produced by 
quantum annealing with $\widehat{Q}$ and the nonlinear power method
is best on the third.
So, overall, a quantum annealing approach is 
best or joint-best on 
five out of the six 
networks.

\section{Discussion}
\label{sec:disc}

We have shown that 
the new objective function 
(\ref{eq:obj1b})
gives a straightforward, parameter-free, method
for (a) judging a core-periphery partition and also (b) determining a partition 
from a real-valued vector of scores.
Moreover, when written in QUBO form the 
resulting discrete maximization 
problem 
is amenable to quantum annealing.
We found that the D-Wave quantum annealer 
could handle QUBOs of this form for networks with 
$\approx 100$ nodes. 
In principle a quantum annealer
is able to find a globally optimum solution.
In practice, of course, as with any physical system 
various sources of noise can affect the performance.
However, we observed that direct application of 
the quantum annealer always produced the best 
or joint-best result
in comparison with 
current heuristic 
core-periphery detection algorithms.
Moreover, a sparsified version of the 
QUBO 
allowed the quantum annealer to be applied 
on networks of size $N = 500$ 
and also gave good results.

Given the likely future performance advancements and increased take-up
of this technology, our work suggests that 
a QUBO/quantum annealing approach has great promise in this 
application area.

We also emphasize that the quantum annealer 
typically delivers multiple  
``samples'' that correspond to 
approximate solutions of the QUBO.
We focused here on the quality of the best sample out of 
100; that is, the binary 
sample $x$ for which the quadratic form 
in (\ref{eq:quboform}) was maximum.
However, we observed
that the full set of samples typically included 
many different almost-optimal alternatives.
Hence the quantum annealer could also be used to
produce coreness scores or rankings across the nodes by, for example,
counting the frequency with which each node was assigned to the core.

We saw in our experiments that the classical (linear) 
eigenvector methods
did not perform well 
in terms of providing approximate 
solutions to the QUBO
(\ref{eq:quboform}).
Intuitively, these types of spectral approaches 
are closely tied with smooth, least-squares type kernels
\cite{spec_hkk,spectralClusteringTutorial}, and hence 
the relaxed optimization problems that they 
solve
are significantly different from 
(\ref{eq:maxobj}). 
The GenBE method from \cite{rombach2017core} aims at maximizing the same type of quadratic spectral kernel, but constrained to a smaller set of indicator-type vectors. While doing better than the purely-quadratic methods, it still did not perform well in our context.
The $h$-index method from 
\cite{kitsak2010identification}
also performed poorly in these tests---the $k$-core construction 
is likely to be more useful when the core-periphery interactions
are less plentiful.
The nonlinear power method 
from \cite{tudisco2019core}
was more successful.
This method directly 
approximates the maximum in 
(\ref{eq:maxobj}) before relaxing to a real-valued 
problem  
and adding a constraint.
More precisely,  
for 
$y,z \in \RR$ and $\alpha > 1$, consider the softmax function 
\begin{equation}
\mu_{\alpha}(y,z) = \left( |y|^\alpha + |z|^\alpha \right)^{1/\alpha}.
\label{eq:mudef}
\end{equation}
Also define the $p$-sphere 
${\cal S}_p = \{ x \in \RR^N : \| x \|_p = 1 \}$
and let 
$ {\cal S}_p^{+} = {\cal S}_p \cap \RR^N_{+}$.
Then the method in \cite{tudisco2019core} gives a globally convergent
iteration for the unique solution of 
\[
\max_{x \in {\cal S}_p^{+}} 
 \sum_{i=1}^{N} \sum_{j=1}^{N} a_{ij} \mu_{\alpha}(x_i,x_j).
\]
Note that for large $\alpha$ the function 
$\mu_{\alpha}(y,z)$ in (\ref{eq:mudef}) is a good approximation to  $\max\{y,z\}$.
Based on 
our results, it would therefore be of interest to 
design and analyse
a similar 
nonlinear power method 
that applies to the 
new objective function  
(\ref{eq:obj1b})
rather than the 
original version (\ref{eq:maxobj}).

\section*{Acknowledgements} 
The work of CFH 
was supported by the 
Engineering and Physical Sciences Research Council 
UK Quantum Technology Programme under grant EP/M01326X/1.
The work of DJH was supported by
the Engineering and Physical Sciences Research Council 
under grants EP/P020720/1 and EP/V015605/1. The work of FT was  supported by the GSSI-SNS Pro3 grant ``STANDS''.  We thank Mason A.\ Porter for supplying code that implements the GenBE algorithm.

% \bibliographystyle{abbrv}
% \bibliography{references}

\end{document}